\documentclass[aps,prd,amssymb,eqsecnum,showpacs,nofootinbib,floatfix,twocolumn,a4paper]{revtex4-1}

\usepackage{amsmath}
\usepackage{graphicx}

\usepackage{color}

\allowdisplaybreaks

\newcommand{\be}{\begin{equation}}
\newcommand{\ee}{\end{equation}}

\newcommand{\Pf}{{\mathrm{Pf}}}

\newcommand{\md}{\mathrm{d}}
\newcommand{\me}{\mathrm{e}}

\newcommand{\vecr}{\mathbf{r}}
\newcommand{\vecp}{\mathbf{p}}
\newcommand{\vecx}{\mathbf{x}}

\newcommand{\vecre}{\mathbf{r}'}
\newcommand{\vecpe}{\mathbf{p}'}
\newcommand{\pepe}{\mathbf{p}'^2}
\newcommand{\nepe}{\mathbf{n}'\cdot\mathbf{p}'}
\newcommand{\rpe}{\mathbf{r}\cdot\mathbf{p}'}

\newcommand{\hr}{\hat{H}_\textrm{real}^\textrm{nr}}
\newcommand{\hri}{\hat{H}_\textrm{real}^\mathrm{nr\,I}}
\newcommand{\hrii}{\hat{H}_\textrm{real}^\mathrm{nr\,II}}
\newcommand{\avhrii}{\hat{\bar{H}}_\textrm{real}^\mathrm{nr\,II}}
\newcommand{\avhii}{\hat{\bar{H}}^\mathrm{nr\,II}}
\newcommand{\hloc}{\hat{H}_\textrm{$\le$ 4PN}^\textrm{local}}

\newcommand{\hreff}{\hat{H}_\textrm{eff}}
\newcommand{\hrieff}{\hat{H}_\textrm{eff}^\mathrm{I}}
\newcommand{\hriieff}{\hat{H}_\textrm{eff}^\mathrm{II}}
\newcommand{\avhriieff}{\hat{\bar{H}}_\textrm{eff}^\mathrm{II}}

\newcommand{\rL}{\mathcal{L}}
\newcommand{\rG}{\mathcal{G}}
\newcommand{\rE}{\mathcal{E}}
\newcommand{\rF}{\mathcal{F}}

\newcommand{\Q}{\hat Q}

\newcommand{\gE}{{\gamma_\textrm{E}}}
\newcommand{\sphys}{s_\textrm{phys}}

\begin{document}

\title{Fourth post-Newtonian effective one-body dynamics}

\author{Thibault Damour}
\email{damour@ihes.fr}
\affiliation{Institut des Hautes Etudes Scientifiques, 35 route de Chartres, 91440 Bures-sur-Yvette, France}

\author{Piotr Jaranowski}
\email{p.jaranowski@uwb.edu.pl}
\affiliation{Faculty of Physics,
University of Bia{\l}ystok,
Cio{\l}kowskiego 1L, 15--245 Bia{\l}ystok, Poland}

\author{Gerhard Sch\"afer}
\email{gos@tpi.uni-jena.de}
\affiliation{Theoretisch-Physikalisches Institut,
Friedrich-Schiller-Universit\"at Jena,
Max-Wien-Pl.\ 1, 07743 Jena, Germany}

%\date{25 February, 2015}

\begin{abstract}
The conservative dynamics of gravitationally interacting two-point-mass systems has been recently
determined at the fourth post-Newtonian (4PN) approximation [T.~Damour, P.~Jaranowski, and G.~Sch\"afer,
Phys.\ Rev.\ D {\bf 89}, 064058 (2014)], and found to be nonlocal-in-time.  We show how to transcribe this dynamics within
the effective one-body (EOB) formalism.  To achieve this EOB transcription, we develop
a new strategy involving the (infinite-)order-reduction of a nonlocal dynamics to an ordinary action-angle Hamiltonian.
Our final, equivalent EOB dynamics comprises two (local) radial potentials, $A(r)$ and $\bar{D}(r)$, and a nongeodesic
mass-shell contribution $Q(r,p_r)$ given by an infinite series of even powers of the radial momentum $p_r$.
Using an effective action technique, we complete our 4PN-level results by deriving two different, higher-order conservative contributions
linked to tail-transported hereditary effects: the 5PN-level EOB logarithmic terms, as well as the 5.5PN-level, half-integral terms.
We compare our improved analytical knowledge to previous, numerical gravitational-self-force computation of precession effects.
\end{abstract}


\maketitle

\section{Introduction}

The impending prospect of detecting the gravitational-wave signals emitted by coalescing binary systems
gives a new incentive for improving our theoretical knowledge of the dynamics of two-body systems in general relativity.
Recent developments have shown that a useful strategy for accurately describing the dynamics of binary systems is to combine,
in a synergetic manner, the information gathered from several different approximation methods: notably, the post-Newtonian (PN) formalism,
the gravitational self-force formalism, full numerical relativity simulations, and the effective one-body (EOB) formalism.
In a recent paper \cite{Damour:2014jta}, we have succeeded in deriving the conservative dynamics of a two-body system
at the 4th post-Newtonian (4PN) approximation [i.e., including fractional corrections of order $\biglb(v^2 / c^2 + GM/(c^2r)\bigrb)^4$
to the Newtonian dynamics]. (Our work was the culmination of previous partial, 4PN-level, investigations
\cite{Blanchet:1987wq,Damour:2009sm,Blanchet:2010zd,Damourlogs,Jaranowski:2012eb,Foffa:2012rn,Jaranowski:2013lca,Bini:2013zaa}.)

The aim of the present work, is to transcribe the Taylor-expanded 4PN dynamics of Ref.\ \cite{Damour:2014jta} within the EOB formalism \cite{Buonanno:1998gg,Buonanno:2000ef,Damour:2000we,Damour:2001tu}. The EOB formalism provides an analytical framework for the description of the relativistic two-body problem which has many useful features: notably, (i) it encompasses natural resummation techniques allowing one to extend the validity of perturbation results up to merger; (ii) it can extract nonperturbative information contained in a few numerical relativity simulations with analytic perturbative information; and (iii) it provides accurate gravitational waveforms corresponding to the full coalescence process from early inspiral to ringdown. (For a sample of recent EOB results see Refs.\ \cite{Damour:2012ky,Pan:2013rra,Taracchini:2013rva,Damour:2014afa,Bernuzzi:2014owa}.)
Until now, the EOB description of the two-body dynamics is fully known only at the 3PN level \cite{Damour:2000we}, though some parts of the EOB description are known to higher PN accuracies. For instance, the main EOB radial potential $A(r)$ is analytically known, to linear order in the mass ratio, up to the 9.5PN level \cite{Bini:2014nfa,Bini:2015bla}, while the logarithmic contributions to the secondary EOB radial potential $\bar D(r)$ are known up to the 5PN level \cite{Blanchet:2010zd,Barack:2010ny,Barausse:2011dq}.

\section{EOB reminder}

To set the stage, let us briefly recall the main features of the EOB formalism.
The basic idea is to map the conservative relative dynamics of a binary system (seen in the center of mass frame)
onto the (equivalent) dynamics of an effective body moving in some effective metric $g_{\mu\nu}^{\rm eff}$
(with additional, nongeodesic, Finsler-type corrections).
This is a general relativistic generalization of the well-known fact
that the Newtonian relative motion of a two-body system is equivalent to the motion of a particle of mass $\mu := m_1 m_2 / (m_1+m_2)$
in the two-body potential $V(r)$. Here, $m_1$ and $m_2$ denote the masses of the binary system.
In the following, we shall also denote
\be
\label{eq1.1}
M := m_1 + m_2, \quad
\nu := \frac\mu M = \frac{m_1m_2}{(m_1+m_2)^2}.
\ee
The full Hamiltonian of the two-body system, in the center of mass frame
($\vecp_1 + \vecp_2 = 0$, $\vecx_{12} := \vecx_1 - \vecx_2$), can be written as
\be
\label{eq1.2}
H [ \vecx_{12} , \vecp_1 ; m_1 , m_2] = Mc^2 + \mu \, \hat{H}^\textrm{nr} [ \vecx_{12} , \vecp_1 ; m_1 , m_2],
\ee
where the reduced, ``nonrelativistic'' Hamiltonian $\hat{H}^\textrm{nr} \equiv {H}^\textrm{nr}/\mu \equiv (H - M c^2)/\mu$ has a PN expansion of the type
\begin{multline}
\label{eq1.3}
\hat{H}^\textrm{nr} [ \vecx_{12} , \vecp_1] = \frac12 \left(\frac{\vecp_1}{\mu} \right)^2 - \frac{GM}{\vert \vecx_{12} \vert} + \frac1{c^2} \, \hat{H}_\textrm{1PN}^\textrm{nr} + \frac1{c^4} \, \hat{H}_\textrm{2PN}^\textrm{nr}
\\[1ex]
+ \frac1{c^6} \, \hat{H}_\textrm{3PN}^\textrm{nr} + \frac1{c^8} \, \hat{H}_\textrm{4PN}^\textrm{nr} + \mathcal{O}\left( \frac1{c^{10}} \right).
\end{multline}
We wish to map the ``real'' Hamiltonian (\ref{eq1.2}) onto some ``effective'' Hamiltonian $H_\textrm{eff} [\vecr' , \vecp']$ describing a relativistic dynamics, with a general mass-shell condition of the type
\be
\label{eq1.4}
\mu^2 c^4 + g_{\rm eff}^{\mu\nu} (x') \, p'_{\mu} p'_{\nu} + Q (x',p') = 0,
\ee
where
\begin{multline}
\label{eq1.5}
Q(x',p') = Q_4^{\mu_1\mu_2\mu_3\mu_4} (x') \, p'_{\mu_1} p'_{\mu_2} p'_{\mu_3} p'_{\mu_4} 
\\[1ex]
+ Q_6^{\mu_1\mu_2\mu_3\mu_4\mu_5\mu_6} (x') \, p'_{\mu_1} p'_{\mu_2} p'_{\mu_3} p'_{\mu_4} p'_{\mu_5} p'_{\mu_6} + \cdots
\end{multline}
denotes contributions which are at least quartic in momenta. As argued in \cite{Damour:2000we} at the 3PN level,
and as we shall show below at the 4PN level, one can reduce the $p'$-dependence of $Q$ to a dependence on the sole radial momentum $p'_r$.
Then, the mass-shell condition (\ref{eq1.4}) is quadratic in the time component $p'_0$,
and $H_{\rm eff} [\vecr' , \vecp']$ denotes the positive-root solution for the effective energy $\rE_{\rm eff} = -p'_0$.

At the 4PN level, the \emph{a priori} unknown functions parametrizing the EOB dynamics comprise:
\begin{enumerate}
\item[(i)] the two functions $A(r')$, $B(r')$ parametrizing a generic spherically symmetric metric (in Schwarzschild-type coordinates)
\begin{multline}
\label{eq1.6}
g_{\mu\nu}^{\rm eff} (x') \, \md x'^{\mu} \md x'^{\nu} = -A(r') \, c^2 \md t'^2 
\\[1ex]
+ B(r') \, \md r'^2 + r'^2 (\md\theta'^2 + \sin^2 \theta' \md\varphi'^2)
\end{multline}
[henceforth, we work with the function $ {\bar D}(r') := (A(r') B(r'))^{-1}$ instead of  $B(r')$];
\item[(ii)] the functions $Q^{rrrr}_4$, $Q_6^{rrrrrr}$, $\ldots$ parametrizing the various contributions to $Q$, Eq.\ (\ref{eq1.5}); and
\item[(iii)] an energy-map function $\rE_{\rm eff} = f(H)$ relating the effective energy $\rE_{\rm eff} = H_{\rm eff}$ to the real energy $\rE_{\rm real} = H$ of the two-body system.
\end{enumerate}

Following \cite{Buonanno:1998gg,Damour:2000we}, we \emph{a priori} allow for a general energy map of the type
\begin{align}
\label{eq1.7}
\rE_{\rm eff} &= \mu \, c^2 + H^{\rm nr} \bigg( 1 + \alpha_1 \, \frac{H^{\rm nr}}{\mu \, c^2} + \alpha_2 \left(\frac{H^{\rm nr}}{\mu \, c^2} \right)^2 
\nonumber\\[1ex]&\qquad
+ \alpha_3 \left(\frac{H^{\rm nr}}{\mu \, c^2} \right)^3 + \alpha_4 \left(\frac{H^{\rm nr}}{\mu \, c^2} \right)^4 + \cdots \bigg),
\end{align}
where the coefficient $\alpha_4$ parametrizes a possible 4PN-level contribution to the energy map $f$.

The energy map $f$ is determined by the requirement that Eq.\ (\ref{eq1.7}) correctly relates the real Hamiltonian (\ref{eq1.2}), (\ref{eq1.3}) to the EOB mass-shell condition (\ref{eq1.4}), (\ref{eq1.5}).
At the 2PN level, it was found that \cite{Buonanno:1998gg}
\be
\label{eq1.8}
\alpha_1 = \frac\nu2 , \quad \alpha_2 = 0 .
\ee
At the 3PN level, it was further found that \cite{Damour:2000we}
\be
\label{eq1.9}
\alpha_3 = 0 .
\ee
One of the new results of the present 4PN-level work has been to find that
\be
\label{eq1.10}
\alpha_4 = 0 .
\ee
This means that the simple energy map found at lower PN levels remains valid at the 4PN level. This energy map can be written as
\be
\label{eq1.11}
\frac{\rE_{\rm eff}}{\mu c^2} \equiv \frac{H_{\rm eff}}{\mu c^2} = \frac{H^2 - m_1^2 \, c^4 - m_2^2 \, c^4}{2 \, m_1 \, m_2 \, c^4},
\ee
or as
\be
\label{eq1.12}
H = M c^2 \, \sqrt{1+2\nu \left( \frac{H_{\rm eff}}{\mu \, c^2} - 1 \right)}.
\ee

\section{Strategy}

After having recalled the basic building blocks of the EOB formalism,
let us outline the strategy we shall use to transcribe the 4PN Hamiltonian of Ref.\ \cite{Damour:2014jta} within the EOB formalism.
The need for a special strategy arises from the main new conceptual feature of the 4PN-level Hamiltonian:
we found in Ref.\ \cite{Damour:2014jta} that, contrary to previous PN levels which led to ordinary (instantaneous) Hamiltonians,
the 4PN conservative dynamics involves both local-in-time interaction terms (described by a usual Hamiltonian)
and a specific (time-symmetric) {\it nonlocal-in-time} interaction.

To explicitly describe this structure, it is convenient to henceforth replace the center-of-mass-frame variables
$\vecp_1=-\vecp_2$ and $\vecx_{12}=\vecx_1-\vecx_2$ used in Eqs.\ (\ref{eq1.2}) and (\ref{eq1.3}) above by the following rescaled variables:
\be
\label{redvar}
\vecr := \frac{\vecx_{12}}{GM},\quad
\vecp := \frac{\vecp_1}{\mu} = -\frac{\vecp_2}{\mu}.
\ee
Note that, in terms of these variables, the Newtonian-level, reduced Hamiltonian takes the simplified form
\be
\label{eq2.2}
\hat H^{\rm nr}_0 (\vecr , \vecp) = \frac12 \, \vecp^2 - \frac1r .
\ee

The (nonrelativistic, reduced) 4PN-accurate Hamiltonian can be decomposed in two parts:
\be
\label{hr1+2}
\hat H^{\rm nr} [\vecr,\vecp] = \hat H^{\rm nr \, I}(\vecr,\vecp;s) + \hat H^{\rm nr \, II}[\vecr,\vecp;s],
\ee
where the first part is local in time while the second part is nonlocal in time.
[We use brackets, $H[\vecr,\vecp]$, instead of round parentheses, $H(\vecr,\vecp)$,
when we wish to emphasize a nonlocal  functional dependence on the phase-space variables.]
More precisely, the Hamiltonian $\hri(\vecr,\vecp;s)$ is a \emph{function} of phase-space variables $(\vecr,\vecp)$ of the form 
\be
\label{eq2.3bis}
\hri(\vecr,\vecp;s) = \hloc(\vecr,\vecp) + F(\vecr,\vecp)\ln\frac{r}{s},
\ee
where the Hamiltonian $\hloc$ is defined in Sec.\ V of \cite{Damour:2014jta}
and the function $F$ is defined in Eq.\ (3.8) there;
$s$ is a reduced scale with dimension of 1/velocity$^2$,
\be
s := \frac{\sphys}{GM},
\ee
where $\sphys$ is a scale with dimension of a length.
The Hamiltonian $\hrii[\vecr,\vecp;s]$ is a \emph{functional} of phase-space trajectories $(\vecr(t),\vecp(t))$,
\begin{multline}
\label{hrii}
\hrii[\vecr(t'),\vecp(t');t,s] = -\frac{1}{5}\frac{G^2}{\nu c^8} \dddot{I}_{\!ij}(t)
\\[1ex]
\times \Pf_{2\sphys/c} \int_{-\infty}^{+\infty} \frac{\md \tau}{\vert \tau \vert} \dddot{I}_{\!ij}(t+\tau),
\end{multline}
where $\mathrm{Pf}_T$ is a Hadamard partie finie with time scale
$T:=2s_\textrm{phys}/c$ [see Eq.\ (4.2) in [1] for the definition]
and where $\dddot{I}_{\!ij}$ denotes a third time derivative of the Newtonian quadrupole moment $I_{ij}$ of the binary system,
\be
I_{ij} := \underset{a}{\sum} \, m_a \left(x_a^i \, x_a^j - \frac13 \, \delta^{ij} \, \mathbf{x}_a^2 \right).
\ee
In terms of the reduced variables \eqref{redvar} it reads
\be
\label{eq1.7J}
I_{ij} = (GM)^2 \mu \left( r^i \, r^j - \frac13 \,  \mathbf{r}^2 \, \delta^{ij} \right).
\ee
Note that the {\it nonlocal} Hamiltonian $\hat H^{\rm nr \, II}[\vecr,\vecp;s]$ slightly differs
from what was defined in \cite{Damour:2014jta} as being the ``nonlocal'' part of the Hamiltonian.
Indeed, there, the 4PN-level nonlocal piece of $H$ was defined by taking as regularization scale in the partie finie operation entering Eq.\ (\ref{hrii})
the length $2\,r_{12}/c$ instead of the $2 \, s_{\rm phys} / c$ appearing in (\ref{hrii}).
As a consequence of this difference, the arbitrary scale $s_{\rm phys}$ enters both parts,
$\hat H^{\rm nr \, I}$ and $\hat H^{\rm nr \, II}$, of $\hat H^{\rm nr}$ though it cancels out in the total Hamiltonian.
Below, we shall separately transcribe each part, $\hat H^{\rm nr \, I}$ and $\hat H^{\rm nr \, II}$, into corresponding parts of the EOB formalism.
We will check at the end that $s_{\rm phys}$ drops out of the final EOB results.

The various techniques used in previous EOB works \cite{Buonanno:1998gg,Damour:2000we} can be directly applied to transform the {\it local} part $\hat H^{\rm nr \, I} (\vecr,\vecp)$ of the 4PN-level Hamiltonian into its corresponding 4PN-level EOB counterparts, $A^{\rm I}_{\leq 4{\rm PN}}$, $B^{\rm I}_{\leq 4{\rm PN}}$, and $Q^{\rm I}_{\leq 4{\rm PN}}$.

By contrast, the matching of the {\it nonlocal} part $\hat H^{\rm nr \, II} [\vecr , \vecp]$ cannot be realized by employing canonical transformations of the usual, {\it local} kind. Here, there are two possibilities for incorporating the physics of $\hat H^{\rm nr \, II} [\vecr , \vecp]$ in an EOB description.
The first possibility would be to introduce some nonlocality in the EOB building blocks $g_{\mu\nu}^{\rm eff}$, $Q$.
[For instance, this could be done simply by transcribing the 4PN piece $\hat H^{\rm nr \, II}=\mathcal{O}(1/c^8)$ into a corresponding nonlocal contribution $Q^{\rm II}=\mathcal{O}(1/c^8)$ to $Q$.] However, there is a useful second possibility when focusing on the dynamics of gravitationally bound systems. [We recall that, currently, the main application of the EOB formalism concerns the description of quasicircular, inspiralling motions (see, however, \cite{Damour:2014afa} for an application to scattering motions).] In that case, it happens that one can formally replace the nonlocality of $\hat H^{\rm nr \, II}$ by local contributions $A^{\rm II}$, $B^{\rm II}$ to the effective metric $g_{\mu\nu}^{\rm eff}$, together with an infinite string of local contributions $Q^{\rm II}$ to $Q$ of the  form
\be
\label{QIIfull}
Q^{\rm II} = \sum_{n=2}^{\infty} Q_{2n}(r') \,  (p'_r)^{2n} .
\ee
In order to do so, several techniques can be used, and combined in various ways. In particular, one could discuss the generalization of the usual, local canonical transformations, to nonlocal ones, with generating functions involving time integrals.
However, it is possible to bypass the explicit use of nonlocal canonical transformations by a suitable combination of other techniques,
as we shall now explain.

Let us first note that $\hat H^{\rm nr \, II}$ being of order $\mathcal{O}(1/c^8)$ can be treated as a first-order effect on top of an ordinary (local) dynamics. Moreover, as we can neglect any cross effect between the various post-Newtonian contributions to $\hat H^{\rm nr  \, I} = \hat H_{\rm Newton}^{\rm nr} + \mathcal{O}(1/c^2)$ and $\hat H^{\rm nr \, II} = \mathcal{O}(1/c^8)$, it is enough to consider $\hat H^{\rm nr \, II}$ as a first-order addition to the Newtonian Hamiltonian, Eq.\ (\ref{eq2.2}). Let us then consider, in some generality, a dynamics of the form
\be
\label{H0H1}
H [\vecr,\vecp] = H_0 (\vecr,\vecp) + \varepsilon \, H_1 [\vecr,\vecp],
\ee
where $H_0 (\vecr,\vecp) = \frac12 \, \vecp^2 - \frac1r$ and where $H_1[\vecr,\vecp]$ is either a higher-order Hamiltonian,
involving a certain number of derivatives of the phase-space variables,
or a nonlocal-in-time Hamiltonian involving integrals over time-separated phase-space variables [such as Eq.\ (\ref{hrii})].
One can formally consider that the second, fully nonlocal, case is equivalent to including an infinite number of time derivatives:
$H_1(\vecr(t),\vecp(t),\ldots,\vecr^{(n)}(t),\vecp^{(n)}(t),\ldots)$.
The first case (involving a finite number of time derivatives) has been abundantly treated in the literature,
in particular within the context of the general relativistic two-body problem
where higher-order Lagrangians and Hamiltonians naturally occur beyond the 1PN level \cite{Schafer:1984mr,Damour:1985mt,Damour:1990jh,Damour:1999cr}.
Of most direct relevance here is the work of Ref.\ \cite{Damour:1999cr}
which considered the order reduction of the higher-order Hamiltonian $H_1 (\vecr , \vecp , \dot\vecr , \dot\vecp)$
arising at the 3PN level in Arnowitt-Deser-Misner coordinates. Let us recall the crucial point made there.
The replacement, within $\varepsilon H_1 (\vecr , \vecp , \dot\vecr , \dot\vecp)$,
of $\dot\vecr$ and $\dot\vecp$ by the equations-of-motion-related expressions
 $\dot\vecr = \vecp + \delta S_H / \delta \vecp + \mathcal{O}(\varepsilon)$,
$\dot\vecp = - \vecr / r^3 - \delta S_H / \delta \vecr + \mathcal{O}(\varepsilon)$
[where $S_H = \int (\vecp \, {\rm d} \vecr - H {\rm d} t)$ denotes the Hamiltonian action],
yields the {\it identity}
\begin{multline}
\label{H1red}
\varepsilon H_1 (\vecr , \vecp , \dot\vecr , \dot\vecp) = \varepsilon H_1^{\rm red} (\vecr , \vecp)
+ \varepsilon \frac{\partial H_1}{\partial \, \dot\vecr} \, \frac{\delta S_H}{\delta \vecp} 
\\[1ex]
- \varepsilon \frac{\partial H_1}{\partial \, \dot\vecp} \, \frac{\delta S_H}{\delta \vecr} + \mathcal{O}(\varepsilon^2) ,
\end{multline}
where $H_1^{\rm red} (\vecr , \vecp) \equiv H_1 (\vecr , \vecp , \vecp , -\vecr / r^3)$ is the naive order-reduced version of $H_1$
[using the $\mathcal{O}(\varepsilon^0)$ equations of motion]. The extra terms in \eqref{H1red} (proportional to the variational derivatives of the action)
immediately lead to the shifts
\be
\delta \vecr = \varepsilon \frac{\partial H_1}{\partial \, \dot\vecp} , \quad \delta \vecp = - \varepsilon \frac{\partial H_1}{\partial \, \dot\vecr} ,
\ee
of the phase-space variables needed to transform the original (higher-order) dynamics for $\vecr , \vecp$ into a dynamics for the shifted variables $\vecr' = \vecr + \delta \vecr$, $\vecp' = \vecp + \delta \vecp$ described by the {\it ordinary} Hamiltonian
\be
\label{Hred}
H^{\rm red} (\vecr' , \vecp') = H_0 (\vecr' , \vecp') + \varepsilon H_1^{\rm red} (\vecr' , \vecp') + \mathcal{O}(\varepsilon^2) .
\ee

It is easily seen that a similar result holds for a higher-order Hamiltonian involving an arbitrary number of derivatives.
Such a result can be further extended to a nonlocal Hamiltonian involving time-shifted phase-space variables,
say $\vecr (t+\tau)$, $\vecp (t+\tau)$, if one uses some integral representation of $\vecr (t+\tau)$, $\vecp (t+\tau)$
in terms of $\vecr (t)$, $\vecp (t)$ and of the extra terms $\delta S_H / \delta\vecp$, $\delta S_H / \delta\vecr$ appearing in the equations of motion written above.
(As usual, one can neglect ``double zero'' terms, i.e.\ work linearly in the variational derivatives of $S_H$.)

Summarizing: Modulo some (nonlocal) shifts of the phase-space variables
(which, in principle, can be read off from the order reduction procedure itself),
one can reduce a nonlocal dynamics of the form (\ref{H0H1}) to an ordinary (local) dynamics of the form 
$H^{\rm red}  (\vecr,\vecp) = H_0 (\vecr,\vecp) + \varepsilon \, H_1^{\rm red}  (\vecr,\vecp) + \mathcal{O}(\varepsilon^2)$,
in which $H_1^{\rm red}  (\vecr,\vecp) $ is the naive order-reduced version of  $H_1 [\vecr,\vecp]$.

\section{Delaunay (action-angle) implementation of the strategy}

In order to apply this strategy, one needs, however, an explicit way of solving the zeroth-order equations of motion
[so as to compute $\vecr (t+\tau)$, $\vecp (t+\tau)$ in terms of $\vecr(t)$, $\vecp(t)$]
and of then computing the order-reduced value of the nonlocal Hamiltonian $\hat H^{\rm nr \, II}$, Eq.\ (\ref{hrii}).
As in our case the zeroth-order [$\mathcal{O}(\varepsilon^0)$] equations of motion are the Newtonian equations of motion,
it is convenient to use the (Delaunay) action-angle form of the Newtonian-level motion.
Let us recall it (using essentially the notation of the book \cite{BC}).
It is enough to consider the planar case.
In that case the action-angle variables are $(\rL , \ell ; \rG , g)$.
Here, the action variables $\rL$, $\rG$ are related to the usual Keplerian variables $a$ (semimajor axis) and $e$ (eccentricity) via
\be
\rL := \sqrt{a},
\quad
\rG := \sqrt{a(1-e^2)},
\ee
or
\be
\label{ae-vs-LG}
a = \rL^2,
\quad
e = \sqrt{1 - \left(\frac{\rG}{\rL}\right)^2}.
\ee

Note that we work here and below with the rescaled variables (\ref{redvar}).
In particular, $a$ denotes the rescaled semimajor axis $a:=a_{\rm phys}/(GM)$,
and the time variable corresponding to the variables $(\rL , \ell ; \rG , g)$, and the corresponding Hamiltonian,
\be
\hat H_0^{\rm nr} = \frac12 \, \vecp^2 - \frac1r = - \frac1{2 \rL^2} ,
\ee
is $\hat t:=t_{\rm phys}/(GM)$.

$\rL$ is conjugate to the ``mean anomaly'' $\ell$, while $\rG$ is conjugate to the argument of the periastron $g=\omega$.
The explicit expressions of the Cartesian coordinates $(x,y)$ of a Newtonian motion in terms of action-angle variables are given by
\begin{subequations}
\begin{align}
\label{eq4.4}
x (\rL , \ell ; \rG , g) &= x_0\cos g - y_0 \sin g,
\\[2ex]
\label{eq4.5}
y (\rL , \ell ; \rG , g) &= x_0\sin g + y_0 \cos g,
\\[2ex]
\label{eq4.6}
x_0 &= a (\cos u - e),
\\[2ex]
\label{eq4.7}
y_0 &= a \sqrt{1-e^2} \sin u,
\end{align}
\end{subequations}
where the ``eccentric anomaly'' $u$ is the function of $\ell$ and $e$ defined by solving Kepler's equation,
\be
\label{Keq}
u - e \sin u = \ell .
\ee
In the equations above, $a$ and $e$ are supposed to be expressed in terms of $\rL$ and $\rG$ using Eq.\ (\ref{ae-vs-LG}).
The solution of Kepler's equation Eq.\ (\ref{Keq}), i.e.\ the explicit expression of $u$ in terms of $\ell$, $\rL$ and $\rG$,
can be written in terms of Bessel functions:
\be
\label{u(l)}
u = \ell + \sum_{n=1}^\infty \frac{2}{n} J_n(ne) \sin(n\ell).
\ee
Note also the following Bessel-Fourier expansions of $\cos u$ and $\sin u$ [which directly enter $(x_0 , y_0)$ and thereby $(x,y)$]
\begin{subequations}
\begin{align}
\label{eq4.10}
\cos u &= -\frac e2 + \sum_{n=1}^{\infty} \frac1n [J_{n-1} (ne) - J_{n+1} (ne)] \cos(n\ell),
\\[2ex]
\label{eq4.11}
\sin u &= \sum_{n=1}^{\infty} \frac1n [J_{n-1} (ne) + J_{n+1} (ne)] \sin(n\ell).
\end{align}
\end{subequations}
For completeness, we also recall the expressions involving the ``true anomaly'' $f$ (polar angle from the periastron) and the radius vector $r$:
\begin{subequations}
\begin{align}
\label{eq4.12}
r = a (1-e \cos u) &= \frac{a (1-e^2)}{1+e \cos f},
\\[2ex]
\frac{x_0}r = \cos f &= \frac{\cos u -e}{1-e \cos u},
\\[2ex]
\frac{y_0}r = \sin f &= \frac{\sqrt{1-e^2} \sin u}{1-e \cos u} .
\end{align}
\end{subequations}

The above expressions allow one to easily evaluate the expansions of $x(\ell)$, $y(\ell)$,
and therefrom the components of the quadrupole tensor $I_{ij} (\ell)$, Eq.\ (\ref{eq1.7J}), as power series in $e$, and Fourier series in $\ell$.
One could also have used the known Bessel-Fourier expansions of the components $I_{ij} (\ell)$ \cite{Peters:1963ux,Blanchet:1989cu,Arun:2007rg}.

Let us then consider the expression
\be
\label{calF}
\mathcal{F}(t,\tau) := \dddot{I}_{\!ij}(t)\dddot{I}_{\!ij}(t+\tau),
\ee
which enters the nonlocal-in-time piece \eqref{hrii} of the Hamiltonian.
In order to evaluate the order-reduced value of $\mathcal{F}(t,\tau)$ we need to use the equations of motion,
both for computing the third time derivatives of $I_{ij}$,
and for expressing the phase-space variables at time $t+\tau$ in terms of the phase-space variables at time $t$.
This is quite easy to do in action-angle variables because the zeroth-order equations of motion
following from the Hamiltonian $H_0 (\rL) = -1 / (2 \rL^2)$ are simply
\begin{subequations}
\begin{align}
\frac{{\rm d}\ell}{{\rm d} \hat t} &= \frac{\partial H_0}{\partial \rL} = \frac1{\rL^3} \equiv \Omega (\rL) , \quad
\frac{{\rm d} g}{{\rm d} \hat t} = \frac{\partial H_0}{\partial \rG} = 0,
\\[3ex]
\frac{{\rm d}\rL}{{\rm d} \hat t} &= - \frac{\partial H_0}{\partial \ell} = 0 , \quad
\frac{{\rm d} \rG}{{\rm d}\hat t} = - \frac{\partial H_0}{\partial g} = 0 .
\end{align}
\end{subequations}
Here, we recall that $\hat t=t/(GM)$ is a rescaled time.
We have introduced the notation $\Omega (\rL) \equiv \rL^{-3}$ for the correspondingly rescaled ($\hat t$-time) Newtonian (anomalistic) orbital frequency:
$\Omega = G M \Omega_{\rm phys}$
(it satisfies the rescaled Kepler law: $\Omega = a^{-3/2}$).
The fact that $g$, $\rL$ and $\rG$ are constant, and that $\ell$ varies linearly with time
makes it easy to compute $\dddot I_{\!ij} (t+\tau)$ in terms of the values of $(\ell,g,\rL,\rG)$ at time $t$.
Namely it suffices to use (denoting by a prime the values at time $t' \equiv t+\tau$)
\be
\ell' \equiv \ell (t+\tau) = \ell(t) + \Omega (\rL) \hat\tau,
\ee
where $\hat\tau \equiv \tau/(GM)$, together with $g' = g$, $\rL' = \rL$, and $\rG' = \rG$.
Finally, the order-reduced value of ${\mathcal F} (t,\tau)$ is given by
(using $\md/\md\hat t = \Omega\,\md/\md\ell$)
\be
\label{calF2}
\mathcal{F}(\ell,\hat\tau) = \bigg(\frac{\Omega (\rL)}{GM}\bigg)^6 \,
\frac{\md^3 I_{ij}}{\md\ell^3}(\ell) \frac{\md^3 I_{ij}}{\md\ell^3}(\ell+\Omega (\rL) \hat\tau).
\ee
Inserting the expansion of $I_{ij} (\ell)$ in powers of $e$ and in trigonometric functions of $\ell$ and $g$
yields $\rF$ in the form of a series of monomials of the type
\be
\label{expF}
\rF(\ell,\hat\tau) = \sum_{n_1 , n_2 , \pm n_3} C_{n_1 n_2 n_3}^{\pm} \, e^{n_1} \cos (n_2 \, \ell \pm n_3 \, \Omega \, \hat\tau),
\ee
where $n_1$, $n_2$, $n_3$ are natural integers.
(Because of rotational invariance, and of the result $g' = g$, there is no dependence of $\rF$ on $g$.)

All the terms in the expansion (\ref{expF}) containing a nonzero value of $n_2$ will,
after integrating over $\hat\tau$ with the measure ${\rm d} \hat\tau / \vert \hat\tau\vert$ as indicated in Eq.\ (\ref{hrii}),
generate a corresponding contribution to $\hat H^{\rm nr \, II}$ which varies with $\ell$ proportionally to $\cos (n_2 \, \ell)$.
At this stage, we appeal to the usual Delaunay technique: any term of the type $A(\rL) \cos (n \ell)$ in a first-order Hamiltonian perturbation $\varepsilon H_1 (\rL , \ell)$
can be eliminated by a canonical transformation with generating function of the type $\varepsilon g(\rL,\ell) = \varepsilon B(\rL) \sin (n\ell)$. Indeed,
\begin{align}
\delta_g H_1 = \{ H_0 (\rL) , g \} &= - \frac{\partial H_0 (\rL)}{\partial \rL} \, \frac{\partial g}{\partial \ell} 
\nonumber\\[2ex]
&= - n \, \Omega (\rL) \, B(\rL) \cos (n\ell),
\end{align}
so that the choice $B = A/(n \, \Omega)$ eliminates the term $A \cos (n\ell)$ in $H_1$.

This shows that all the periodically varying terms (with $n_2 \ne 0$) in $\rF$, Eq.\ (\ref{expF}),
can be eliminated by a canonical transformation.
This proves that one can finally further simplify the (order-reduced) second (nonlocal) part of the 4PN Hamiltonian
by replacing it by its $\ell$-averaged value,
\be
\label{avhrii}
\avhrii(\rL,\rG;s) := \frac{1}{2\pi}\int_0^{2\pi}\md\ell\,\hrii[\vecr,\vecp;s],
\ee
i.e.,
\be
\label{avHii}
\avhii(\rL,\rG;s) = -\frac15 \, \frac{G^2}{\nu c^8} \, {\rm Pf}_{2s/c}
\int_{-\infty}^{+\infty} \frac{{\rm d} \hat\tau}{\vert \hat\tau \vert} \, \bar\rF ,
\ee
where $\bar\rF$ denotes the $\ell$-average of $\rF (\ell , \hat\tau)$
[which is simply obtained by dropping all the terms with $n_2 \ne 0$ in the expansion (\ref{expF})].
This procedure yields an averaged Hamiltonian $\hat{\bar{H}}^\mathrm{nr\,II}$ which depends only on $\rL$, $\rG$ (and $s$),
and which is given as an expansion in powers of $e$. Because of the averaging the latter expansion contains only even powers of $e$.

The final step of our strategy will be to match the latter Hamiltonian to a corresponding piece in the EOB Hamiltonian.
This matching is naturally done by performing an analog Delaunay reduction of the corresponding piece in the EOB Hamiltonian,
say $\hat H_{\rm EOB}^{\rm II}$. To start with, $\hat H_{\rm EOB}^{\rm II}$ is given, from the mass-shell condition (\ref{eq1.4}), (\ref{eq1.5}),
as a function of $\vecp'^2$, $1/r'$ and $p'^2_r$, and contains an \emph{a priori} infinite string of powers of $p'^2_r$:
$Q_4 (r') \, p'^4_r + Q_6 (r') \, p'^6_r + \cdots$.
As $p'_r$ is of order of the eccentricity, and as we noticed that $\hat{\bar{H}}^\mathrm{nr\,II}$ contains only even powers of $e$,
we see that (as anticipated) it is enough to include in $Q$ only terms even in $p'_r$.

Having explained beforehand our strategy,
we shall successively implement its various steps in the following sections.

\begin{widetext}
\section{Split of the effective EOB Hamiltonian}

Henceforth we shall set $c=1$ for simplicity.  Let us recall that the reduced effective-one-body  4PN-accurate Hamiltonian
(expressed in the coordinates $\vecre$, $\vecpe$ of the effective dynamics), i.e.\ the solution of the EOB
mass-shell condition, takes the following explicit form [in which $\bar D := (A B)^{-1}$, $\Q := Q/\mu^2$]:
%\begin{widetext}
\begin{align}
\label{hreff}
\hat{H}_\textrm{eff}(\vecre,\vecpe) := \frac{H_\textrm{eff}(\vecre,\vecpe)}{\mu }
= \sqrt{A(r')\bigg(1+\pepe+\Big(A(r')\bar{D}(r')-1\Big)(\nepe)^2+\Q(\vecre,\vecpe)\bigg)}.
\end{align}
\end{widetext}

To the (local versus nonlocal) split  Eq.\ (\ref{hr1+2}) of the  two-body Hamiltonian, there corresponds a (4PN-accurate)  split of the various 
building blocks $A$, $\bar{D}$, and $\Q$  entering the effective Hamiltonian of the form
\begin{subequations}
\label{eobAD}
\begin{align}
A(r') &= A^\textrm{I}(r') + A^\textrm{II}(r'),
\\[1ex]
\label{eobAD2}
A^\textrm{I}(r') &= 1 -  \frac{2}{r'} + \frac{2\nu}{r'^3} + \frac{a_4}{r'^4}
+ \frac{a_\textrm{5,c}^\textrm{I} + a_\textrm{5,ln}^\textrm{I}\ln r'}{r'^5},
\\[1ex]
A^\textrm{II}(r') &= \frac{a_\textrm{5,c}^\textrm{II}+a_\textrm{5,ln}^\textrm{II}\ln r'}{r'^5},
\\[2ex]
\bar{D}(r') &= \bar{D}^\textrm{I}(r') + \bar{D}^\textrm{II}(r'),
\\[1ex]
\label{eobAD5}
\bar{D}^\textrm{I}(r') &= 1 + \frac{6\nu}{r'^2} + \frac{\bar{d}_3}{r'^3}
+ \frac{\bar{d}_\textrm{4,c}^\textrm{\,I} + \bar{d}_\textrm{4,ln}^\textrm{\,I}\ln r'}{r'^4},
\\[1ex]
\bar{D}^\textrm{II}(r') &= \frac{\bar{d}_\textrm{4,c}^\textrm{\,II} + \bar{d}_\textrm{4,ln}^\textrm{\,II}\ln r'}{r'^4}.
\end{align}
\end{subequations}
In Eqs.\ \eqref{eobAD} we have used the explicit values of the 2PN-accurate coefficients
in the potentials $A^\textrm{I}$ and $\bar{D}^\textrm{I}$,
the two coefficients $a_4$ and $\bar{d}_3$ are at 3PN level, and the remaining eight coefficients
($a_\textrm{5,c}^\textrm{I}$, $a_\textrm{5,ln}^\textrm{I}$,
$a_\textrm{5,c}^\textrm{II}$, $a_\textrm{5,ln}^\textrm{II}$,
$\bar{d}_\textrm{4,c}^\textrm{\,I}$, $\bar{d}_\textrm{4,ln}^\textrm{\,I}$,
$\bar{d}_\textrm{4,c}^\textrm{\,II}$, $\bar{d}_\textrm{4,ln}^\textrm{\,II}$) are at the 4PN level.

The corresponding split of the ``nongeodesic'' term $\Q$ of the effective-one-body Hamiltonian \eqref{hreff}
is taken with the following structure (which will be checked to be adequate):
%\begin{widetext}
\begin{subequations}
\label{eobQ}
\begin{align}
\Q(\vecre,\vecpe) &= \Q^\textrm{I}(\vecre,\vecpe) + \Q^\textrm{II}(\vecre,\vecpe),
\\[1ex]
\label{eobQ2}
\Q^\textrm{I}(\vecre,\vecpe) &= \bigg( \frac{q_{42}}{r'^2}
+ \frac{q_\textrm{43,c}^\textrm{I} + q_\textrm{43,ln}^\textrm{I}\ln r'}{r'^3} \bigg) (\nepe)^4
\nonumber\\[1ex]&\quad
+ \frac{q_\textrm{62,c}^\textrm{I} + q_\textrm{62,ln}^\textrm{I}\ln r'}{r'^2}(\nepe)^6,
\\[1ex]
\Q^\textrm{II}(\vecre,\vecpe) &=
\frac{q_\textrm{43,c}^\textrm{II} + q_\textrm{43,ln}^\textrm{II}\ln r'}{r'^3} (\nepe)^4
\nonumber\\[1ex]&\quad
+ \frac{q_\textrm{62,c}^\textrm{II} + q_\textrm{62,ln}^\textrm{II}\ln r'}{r'^2}(\nepe)^6
\nonumber\\[1ex]&\quad
+ \mathcal{O}\big((\nepe)^8\big).
\end{align}
\end{subequations}
%\end{widetext}
Here the coefficient $q_{42}$ represents the 3PN order, while the eight coefficients
$q_\textrm{43,c}^\textrm{I}$, $q_\textrm{43,ln}^\textrm{I}$,
$q_\textrm{43,c}^\textrm{II}$, $q_\textrm{43,ln}^\textrm{II}$,
$q_\textrm{62,c}^\textrm{I}$, $q_\textrm{62,ln}^\textrm{I}$,
$q_\textrm{62,c}^\textrm{II}$, and $q_\textrm{62,ln}^\textrm{II}$
are the 4PN level.

In the formulas above, all the 3PN-level coefficients have been determined in our previous work \cite{Damour:2000we}, namely,
\begin{align*}
a_4 &= \left(\frac{94}{3}-\frac{41\pi^2}{32}\right)\nu,
\\[1ex]
\bar{d}_3 &= 52\nu - 6\nu^2,
\\[1ex]
q_{42} &= 2(4-3\nu)\nu.
\end{align*}

When inserting the above I $+$ II split of all the functions $A$, $\bar{D}$, and $\Q$ entering the effective Hamiltonian,
one obtains, after expanding
the right-hand side of Eq.\ \eqref{hreff} into a Taylor series with respect to $1/c^2$ ($\sim \vecpe^2 \sim 1/r'$), the following split of $\hreff$:
\be
\label{hreff1+2}
\hreff(\vecre,\vecpe) = \hrieff(\vecre,\vecpe) + \hriieff(\vecre,\vecpe)
+ \mathcal{O}(\textrm{5PN}).
\ee
Here the first effective Hamiltonian $\hrieff$ is computed from the functions $A^\textrm{I}$, $\bar{D}^\textrm{I}$, and $\Q^\textrm{I}$
(as if $A^\textrm{II}$, $\bar{D}^\textrm{II}$, and $\Q^\textrm{II}$ were equal to zero),
while the second effective Hamiltonian $\hriieff$ is linear in the 4PN-level functions $A^\textrm{II}$, $\bar{D}^\textrm{II}$, and $\Q^\textrm{II}$.
Note that both parts of the effective Hamiltonian are {\it linear} in the 4PN-level contributions to the  functions $A$, $\bar{D}$, and $\Q$.
For instance, the (simpler) second part of the effective Hamiltonian reads (at the 4PN accuracy)
\begin{align}
\hriieff(\vecre,\vecpe) = \frac{1}{2}
\Big(A^\textrm{II}(r') + \bar{D}^\textrm{II}(r')(\nepe)^2
%\nonumber\\
+ \Q^\textrm{II}(\vecre,\vecpe)\Big).
\end{align}

\section{Matching of the first (local) part of the Hamiltonian to the first part of the EOB dynamics}

In this section, we shall consider the matching of the
{\it first} part of the two-body Hamiltonian, i.e.\ the (local) part $\hri$  in Eq.\ (\ref{hr1+2}), to the corresponding first part
of the EOB dynamics, described by the functions $A^\textrm{I}$, $\bar{D}^\textrm{I}$, and $\Q^\textrm{I}$
discussed in the previous section. (Note that this first part involves all the known, lower PN contributions.)
This matching could be done by any of the various techniques used in previous EOB works \cite{Buonanno:1998gg,Damour:2000we}.
Here, we shall use the technique of \cite{Damour:2000we} based on requiring that the EOB phase-space variables $(\vecr',\vecp')$
differ from the original (rescaled) ones $(\vecr,\vecp)$ by a canonical transformation.
It is at this stage that one should \emph{a priori} allow for a general energy map
Eq.\ (\ref{eq1.7}), possibly involving a new, 4PN-level parameter $\alpha_4$, between the real two-body energy and the effective one.
In other words, the two-body/EOB matching should \emph{a priori} be done by writing that
\begin{multline}
(\hat{H}_\textrm{eff}(\vecre,\vecpe))^2 = \bigg( 1 + \hr(\vecr,\vecp)
\Big( 1 + \frac{\nu}{2}\hr(\vecr,\vecp) 
\\
+ \alpha_3(\hr(\vecr,\vecp))^3
+ \alpha_4(\hr(\vecr,\vecp))^4 \Big)\bigg)^2 ,
\end{multline}
and by splitting such a matching equation in two parts, according to our general I $+$ II split.
However, we already mentioned above that one of our results at the 4PN level is that the energy map Eq.\ (\ref{eq1.7}) between the
real two-body energy and the effective one does not need to be changed compared to previous EOB results \cite{Buonanno:1998gg,Damour:2000we},
i.e.\ that we have simply $\alpha_4=0$, Eq.\ (\ref{eq1.10}). 

Matching the part $\hri$ of the two-body Hamiltonian to the part $\hrieff$ of the effective EOB Hamiltonian means
that the equality
\begin{align}
(\hrieff(\vecre,\vecpe))^2 = \bigg( 1 + \hri(\vecr,\vecp)
\Big( 1 + \frac{\nu}{2}\hri(\vecr,\vecp)  \Big) \bigg)^2
\end{align}
holds modulo a canonical transformation between the phase-space variables $(\vecr,\vecp)$, $(\vecre,\vecpe)$
given by some unknown generating function with the symbolic structure
\begin{multline}
G(\vecr,\vecpe) = G_\textrm{$\le$3PN}(\vecr,\vecpe)
+ (\rpe)(1+\ln r)
\\[1ex]
\times\bigg((\pepe)^4 + \frac{1}{r}\big((\pepe)^3+\cdots\big) + \cdots + \frac{1}{r^4} \bigg) ,
\end{multline}
where  $G_\textrm{$\le$3PN}$ is known \cite{Damour:2000we}, and where the omitted coefficients entering the
4PN level are to be found.

We found a \emph{unique} solution to this  4PN-level matching. Let us only give here the results for
the physically most relevant information, i.e.\ the first parts of the various EOB building blocks. They are
\begin{widetext}
\begin{subequations}
\label{match1res}
\begin{align}
& a_\textrm{5,c}^\textrm{I} = \left(-\frac{4237}{60}+\frac{2275\pi^2}{512}-\frac{128}{5}\ln s\right)\nu
+ \left(\frac{41\pi^2}{32}-\frac{221}{6}\right)\nu^2,
\quad
a_\textrm{5,ln}^\textrm{I} = \frac{128}{5}\nu,
\\[1ex]
& {\bar{d}_\textrm{4,c}}^\textrm{\,I} = \left(\frac{7243}{45}-\frac{23761\pi^2}{1536}-\frac{1184}{15}\ln s\right)\nu
+ \left(\frac{123\pi^2}{16}-260\right) \nu^2,
\quad
\bar{d}_\textrm{4,ln}^\textrm{\,I} = \frac{1184}{15}\nu,
\\[1ex]
& q_\textrm{43,c}^\textrm{I} = 20\nu - 83\nu^2 + 10\nu^3,
\quad
q_\textrm{43,ln}^\textrm{I} = 0,
\quad
q_\textrm{62,c}^\textrm{I} = -\frac{9}{5}\nu - \frac{27}{5}\nu^2 + 6\nu^3,
\quad
q_\textrm{62,ln}^\textrm{I} = 0.
\end{align}
\end{subequations}
\end{widetext}

\section{Matching of the second (nonlocal) part of the Hamiltonian to the second part of the EOB dynamics}

We have explained above our strategy for the more subtle nonlocal-in-time second part of the two-body Hamiltonian.
It involves three steps:
(i) one must explicitly compute the $\ell$-averaged, order-reduced nonlocal part of the two-body Hamiltonian  \eqref{hrii};
(ii) one must separately compute the  $\ell$-average of the second part  $\hriieff$ of the effective EOB Hamiltonian;
and then (iii) one must match the two results.

We have indicated above the Bessel-expansion technique allowing one to compute $\avhrii$ to any
preassigned order in eccentricity. This technique can be applied in various ways,
e.g.: (i) by inserting the $e$ expansion of Eq.\ (\ref{u(l)}), i.e.\ (up to $e^5$ for illustration)
\begin{align}
u &= \ell + \left( e - \frac18 \, e^3 + \frac1{192} \, e^5 \right) \sin \ell
\nonumber \\[1ex]&\quad
+  \left( \frac12 \, e^2 - \frac16 \, e^4 \right) \sin 2 \ell
\nonumber \\[1ex]&\quad
+ \left( \frac38 \, e^3 - \frac{27}{128} \, e^5 \right) \sin 3 \ell + \frac13 \, e^4 \sin 4 \ell
\nonumber \\[1ex]&\quad
+ \frac{125}{384} \, e^5 \sin 5 \ell + {\mathcal O}(e^6),
\end{align}
directly into Eqs.\ (\ref{eq4.4})--(\ref{eq4.7}) to compute the $e$-expansion of the quadrupole moment $I_{ij} (\ell , e)$;
or (ii) to use Eqs.\ (\ref{eq4.10})--(\ref{eq4.11}) as intermediate results;
or (iii) to use the Bessel-Fourier expansion of the quadrupole moment \cite{Peters:1963ux,Blanchet:1989cu,Arun:2007rg},
say (where $\me=2.718281828\ldots$ should be distinguished from the eccentricity $e$)
\be
\label{IBessel}
I_{ij} (\ell ; e) = \sum_{p=-\infty}^{+\infty} I_{ij} (p) \, \me^{ip\ell} .
\ee

We used, as a check on the results, several of these techniques, and we have pushed the calculation to the seventh order in $e$.
Using Eq.\ (\ref{IBessel}), we see that the structure of the $\ell$-average $\bar{\mathcal F}$ of ${\mathcal F}(\ell,\hat\tau)$,
Eq.\ (\ref{calF2}), will be 
\begin{align}
\label{Faverage}
\bar{\mathcal F} (\hat\tau) &= \left( \frac{\Omega}{GM} \right)^6
\sum_{p=-\infty}^{+\infty} p^6 \vert I_{ij} (p) \vert^2 \, \me^{ip\Omega \hat\tau}
\nonumber \\[1ex]
&= 2 \left( \frac{\Omega}{GM} \right)^6 \sum_{p=1}^{\infty} p^6 \vert I_{ij} (p) \vert^2 \cos(p\,\Omega\,\hat\tau). 
\end{align}
The computation of the averaged Hamiltonian $\hat{\bar H}^{\rm nr \, II}$, given by Eq.\ (\ref{avHii}), is then reduced to a series of integrals 
(here written in their one-sided forms) of the type [see, e.g., Eq.\ (5.8) in \cite{Damour:2014jta}],
\be
\label{intcos}
\Pf_{T} \int_0^{+\infty} \frac{\md \hat \tau }{ \hat \tau} \cos(\omega \hat\tau)
= -(\gamma_\textrm{E} + \ln(| \omega | T)),
\ee
where $\omega = p \Omega$ is some integer multiple of $\Omega$. 
This yields the series representation
\begin{multline}
\label{eq7.4}
\avhrii(\rL,\rG;s)= \frac45 \frac{G^2}{\nu c^8}  \left( \frac{\Omega}{GM} \right)^6 
\\[1ex]
\times \sum_{p=1}^{\infty} p^6 \vert I_{ij} (p) \vert^2  \ln\left( 2 p \frac{\me^{\gE}\Omega s}{c}\right).
\end{multline}
(Similarly looking, but different, series appear in tail contributions to gravitational wave fluxes \cite{Arun:2007rg}.)
Let us define the dimensionless reduced quadrupole moment
\be
\hat{I}_{ij} := \frac{I_{ij}}{(GM)^2\mu a^2}.
\ee
Then, making use of the Bessel-Fourier expansion of the individual components of the quadrupole moment
[see, e.g., Eqs.\ (A3) in \cite{Arun:2007rg}], one can show that
\begin{widetext}
\begin{align}
\vert \hat{I}_{ij} (p) \vert^2 &= \frac{8}{3e^4p^4}
\Big(3 e^2 \left(1-e^2\right) \big(1+p^2-p^2e^2\big) J_{p-1}^2(e p)
\nonumber\\&\quad
-3 e(1-e^2) \big(2(p+1)^2-p(2 p+3)e^2\big) J_{p-1}(e p) J_p(e p)
\nonumber\\&\quad
+ \left(6(p+1)^2 - 3 (p+1) (5 p+2)e^2  + \left(12 p^2+9 p+1\right)e^4 -3 p^2 e^6 \right) J_p^2(e p)\Big).
\end{align}
Because of the factor $e^4$ in the denominator, this (simplified) expression hides the fact 
that the term of index $p$ is of order $\mathcal{O}(e^{2|p-2|})$.
To see this, one can use the following equivalent (but more complicated) expression:
\begin{multline}
\vert\hat{I}_{ij}(p)\vert^2 = \frac{1}{576} \bigg\{ 18(1-e^2) \Biglb(
5e \big(J_{p+1}(e p)-J_{p-1}(e p)\big)
+ 2(e^2+1) \big(J_{p-2}(ep)-J_{p+2}(e p)\big)
+ e \big(J_{p+3}(e p)-J_{p-3}(ep)\big) \Bigrb)^2
\\[1ex] + \Biglb(
4(9e^2+1) J_p(e p)
- 3e(3e^2+7) \big(J_{p-1}(e p)+J_{p+1}(e p)\big)
+ 6(e^2+1) \big(J_{p-2}(e p)+J_{p+2}(e p)\big)
+ e(e^2-3) \big(J_{p-3}(e p)+J_{p+3}(e p)\big) \Bigrb)^2
\\[1ex] + \Biglb(
4(6e^2-1) J_p(e p)
- 3e(2e^2+3) \big(J_{p-1}(e p)+J_{p+1}(e p)\big)
+ 6 \big(J_{p-2}(e p)+J_{p+2}(e p)\big)
+ e(2 e^2-3) \big(J_{p-3}(e p)+J_{p+3}(e p)\big) \Bigrb)^2
\\[1ex] + \Biglb(
- 4(3e^2+2) J_p(e p)
+ 3e(e^2+4) \big(J_{p-1}(e p)+J_{p+1}(e p)\big)
- 6e^2 \big(J_{p-2}(e p)+J_{p+2}(e p)\big)
+ e^3 \big(J_{p-3}(e p)+J_{p+3}(e p)\big) \Bigrb)^2 \bigg\}.
\end{multline}
The final result for the \emph{averaged} Hamiltonian $\avhrii$ reads
\begin{align}
\avhrii(\rL,\rG;s) &= \frac{\nu}{c^8\rL^{10}} \Bigg(
\frac{64}{5} \bigglb(2\ln 2 + \ln\left(\frac{\me^\gE s}{c\rL^3}\right)\biggrb)
+ \frac{1}{5} \bigglb( \frac{296}{3}\ln2 + 729\ln3
+ \frac{1256}{3}\ln\left(\frac{\me^\gE s}{c\rL^3}\right)\biggrb) e^2
\nonumber\\[1ex]&\qquad
+ \bigglb(\frac{29966}{15}\ln2 - \frac{13851}{20}\ln3
+ 242 \ln\left(\frac{\me^\gE s}{c\rL^3}\right)\biggrb) e^4
\nonumber\\[1ex]&\qquad
+ \bigglb(-\frac{116722}{15}\ln2 + \frac{419661}{320}\ln3 + \frac{1953125}{576}\ln5
+ \frac{1526}{3} \ln\left(\frac{\me^\gE s}{c\rL^3}\right)\biggrb) e^6
+ \mathcal{O}(e^8) \Bigg).
\end{align}
\end{widetext}
We recall that the eccentricity $e$ is considered to be the function of $\rL$ and $\rG$ 
given by Eq.\ \eqref{ae-vs-LG}.

Let us now consider the second step, i.e.\ the $\ell$-average of 
the part $\hriieff$ of the Taylor-expanded Hamiltonian $\hreff$, expressed 
in terms of Delaunay variables $\rL$, $\rG$, $\ell$ and $g$. In view of the explicit
expression of  $\hriieff$ (with the above-given forms of $A^\textrm{II}$, $\bar{D}^\textrm{II}$, and $\Q^\textrm{II}$),
the result depends on the  $\ell$-average of  monomials involving various powers of $1/r'$ and $p'_r$ 
(and sometimes a logarithm of $r'$).  All those computations (which must be done along
the Newtonian motion) can be straightforwardly performed 
by expanding all formulas in the eccentricity $e$ up to the order $e^7$, using the standard 
Keplerian-motion results recalled above. We found
\begin{widetext}
\begin{align}
\avhriieff(\rL,\rG) &:= \frac{1}{2\pi}\int_0^{2\pi}\md\ell\,\hriieff[\vecre,\vecpe]
\nonumber\\[1ex]
&\phantom{:}= \frac{1}{2\rL^{10}} \Bigg(
a_\textrm{5,c}^\textrm{II} + a_\textrm{5,ln}^\textrm{II} \ln(\rL^2)
+ \frac{1}{4} \left( 20 a_\textrm{5,c}^\textrm{II} - 9 a_\textrm{5,ln}^\textrm{II}
+ 2 \bar{d}_\textrm{4,c}^\textrm{\,II} + 2 (10 a_\textrm{5,ln}^\textrm{II}
+ \bar{d}_\textrm{4,ln}^\textrm{\,II})\ln(\rL^2) \right) e^2
\nonumber\\[1ex]&\qquad
+ \left( \frac{1}{8} \bigg(105 a_\textrm{5,c}^\textrm{II}
- \frac{319}{4}a_\textrm{5,ln}^\textrm{II} + 15 \bar{d}_\textrm{4,c}^\textrm{\,II}
- \frac{11}{2}\bar{d}_\textrm{4,ln}^\textrm{\,II} + 3 q_\textrm{43,c}^\textrm{II} \bigg)
+ \frac{3}{8} \big(35 a_\textrm{5,ln}^\textrm{II}
+ 5 \bar{d}_\textrm{4,ln}^\textrm{\,II} + q_\textrm{43,ln}^\textrm{II}\big)\ln(\rL^2)
\right) e^4
\nonumber\\[1ex]&\qquad
+ \bigg( \frac{1}{192} \left(5040 a_\textrm{5,c}^\textrm{II} - 5018a_\textrm{5,ln}^\textrm{II}
+ 840 \bar{d}_\textrm{4,c}^\textrm{\,II} - 533\bar{d}_\textrm{4,ln}^\textrm{\,II}
+ 252 q_\textrm{43,c}^\textrm{II} - 78 q_\textrm{43,ln}^\textrm{II} + 60  q_\textrm{62,c}^\textrm{II}\right)
\nonumber\\[1ex]&\qquad
+ \frac{1}{16} \big(420a_\textrm{5,ln}^\textrm{II} + 70\bar{d}_\textrm{4,ln}^\textrm{\,II}
+ 21 q_\textrm{43,ln}^\textrm{II} + 5 q_\textrm{62,ln}^\textrm{II}\big) \ln(\rL^2) \bigg) e^6
+ \mathcal{O}(e^8) \Bigg).
\end{align}

The last step is to match the contributions of these two averaged dynamics to the real (rather than
effective) Hamiltonian. Here, we do not need to invoke any canonical transformation because the
action variables of the real and effective dynamics are to be identified \cite{Buonanno:1998gg}.
One must still take into account the \emph{a priori} nontrivial energy map relating the real
and effective energies. However, it is easily seen that, as we are here dealing, on both sides, with an
additional contribution of the 4PN level, it is enough to impose the requirement
\be
\avhriieff(\rL,\rG) = \avhrii(\rL,\rG).
\ee
The \emph{unique} result of this matching then gives
\begin{subequations}
\label{match2res}
\begin{align}
& a_\textrm{5,c}^\textrm{II} = \frac{128}{5} (\gamma_\textrm{E} + 2\ln2+ \ln s) \nu,
\quad
a_\textrm{5,ln}^\textrm{II} = -\frac{192}{5}\nu,
\\[1ex]
& {\bar{d}_\textrm{4,c}}^\textrm{\,II} = \left(-\frac{864}{5}+\frac{1184}{15}\gamma_\textrm{E} - \frac{6496}{15}\ln2 + \frac{2916}{5}\ln3
+\frac{1184}{15}\ln s \right)\nu,
\quad
\bar{d}_\textrm{4,ln}^\textrm{\,II} = -\frac{592}{5}\nu,
\\[1ex]
& q_\textrm{43,c}^\textrm{II} = \left(-\frac{5608}{15} + \frac{496256}{45}\ln2 -\frac{33048}{5}\ln3\right)\nu,
\quad
q_\textrm{43,ln}^\textrm{II} = 0,
\\[1ex]
& q_\textrm{62,c}^\textrm{II} = \left(-\frac{4108}{15} - \frac{2358912}{25}\ln2 + \frac{1399437}{50}\ln3 + \frac{390625}{18}\ln5\right)\nu,
\quad
q_\textrm{62,ln}^\textrm{II} = 0.
\end{align}
\end{subequations}

\section{Final results for the 4PN-accurate EOB dynamics}

Adding the results \eqref{match1res} and \eqref{match2res} of the matching procedures of the two
parts of the dynamics, finally yields the 4PN-accurate form of the EOB functions $A$, $\bar{D}$, and $\Q$.
They read (we use here $u\equiv1/r'$)
\begin{subequations}
\label{eqs8.1}
\begin{align}
A(u) &= 1 - 2 u + 2 \nu \,u^3 + \left(\frac{94}{3}-\frac{41\pi^2}{32}\right) \nu \,u^4
\nonumber\\[1ex]&\quad + \bigglb(
\left( \frac{2275\pi^2}{512} - \frac{4237}{60} + \frac{128}{5}\gE + \frac{256}{5}\ln2 \right) \nu
+ \left( \frac{41\pi^2}{32} - \frac{221}{6} \right)\nu^2
+ \frac{64}{5}\nu\,\ln u \biggrb) u^5,
\\[2ex]
\bar{D}(u) &= 1 + 6\nu\,u^2 + \left(52\nu - 6\nu^2\right) u^3
\nonumber\\[1ex]&\quad + \bigglb(
\left( -\frac{533}{45} - \frac{23761\pi^2}{1536} + \frac{1184}{15}\gE - \frac{6496}{15}\ln2 + \frac{2916}{5}\ln3 \right)\nu
+ \left(\frac{123\pi^2}{16} - 260\right)\nu^2 + \frac{592}{15}\nu\ln u \biggrb) u^4,
\\[2ex]
\Q(\vecre,\vecpe) &= \bigglb( 2(4-3\nu )\nu\,u^2
+ \left( \left( -\frac{5308}{15} + \frac{496256}{45}\ln2 - \frac{33048}{5}\ln3 \right) \nu
- 83 \nu^2 + 10 \nu^3 \right) u^3 \biggrb) (\nepe)^4
\nonumber\\[1ex]&\quad
+ \bigglb( \left(-\frac{827}{3} - \frac{2358912}{25}\ln2 + \frac{1399437}{50}\ln3
+ \frac{390625}{18}\ln5 \right)\nu - \frac{27}{5}\nu^2 + 6\nu^3 \biggrb)u^2\,(\nepe)^6 + \mathcal{O}[ \nu u  (\nepe)^8].
\end{align}
\end{subequations}
\end{widetext}

Various comments are in order. Let us first emphasize that,
while the arbitrary scale $s$ entered the separate pieces I and II,
it has (as expected) dropped out of the final results.
Second, we note that the 4PN-accurate result for $A(u)$ is not new,
but confirms the result first obtained in \cite{Bini:2013zaa}.
The new results with this paper concern the 4PN contributions to the EOB potentials  $\bar{D}(u) $ and $\Q(u, p'_r)$,
except for the 4PN logarithmic contribution to  $\bar{D}(u) $ which was previously derived in 
\cite{Blanchet:2010zd,Damourlogs,LeTiec:2011ab,Barausse:2011dq}.
For completeness, as the derivation used in  \cite{Damourlogs} was never published,
we indicate its main steps in the following section.
Concerning the nonlogarithmic 4PN term in $\bar{D}(u) $ that is {\it linear } in $\nu$, i.e.\ the term of order $\nu u^4$, it was emphasized
in \cite{Damour:2009sm} that it was computable via gravitational self-force calculations, and a numerical estimate of its value
[combined with some $A(u)$-related contributions] had been previously obtained in the work of Ref.\ \cite{Barack:2010ny} (with further processing in 
Ref.\ \cite{Barausse:2011dq}).  For comparison, let us start by noticing that the
numerical values of the coefficients of the 4PN-accurate function $\bar{D}(u)$ analytically obtained by us read
\begin{align}
\bar{D}(u) = 1 &+ 6 \nu u^2 + (52 \nu - 6 \nu^2) u^3
\nonumber\\
&+ (221.572\,\nu - 184.127\,\nu^2 + 39.4667\,\nu\ln u) u^4.
\end{align}

Let us compare the analytical value of the coefficient of the term of order $\nu u^4$, i.e.\ $ \bar{d}_\textrm{4,c}^{\,1 \, {\rm ana}} \approx 221.572$, 
to the numerical estimate deduced from the 
gravitational self-force calculations of  \cite{Barack:2010ny}.  From the further work of  \cite{Barausse:2011dq} (which also made use
of numerical results of \cite{LeTiec:2011ab}) it was $ \bar{d}_\textrm{4,c}^{\,1 \, {\rm num}} = 226^{+7}_{-4}$, which agrees with 
$ \bar{d}_\textrm{4,c}^{\,1 \, {\rm ana}} = 221.5719912$, within, essentially, the 
 one sigma level. Let us also compare to the quantity which was directly numerically evaluated in \cite{Barack:2010ny}, namely the
following combination of  3PN- and 4PN-level coefficients
of the functions $A(u)$ and $\bar{D}(u)$:
\be
\label{BDS}
\rho^\textrm{c}_4 = -\frac{27}{4} - 7 a_4 + 10 a_\textrm{5,c}^1 - 6 \bar{d}_\textrm{3,c}^{\,1}
+ \bar{d}_\textrm{4,c}^{\,1} + \frac{9}{2} a_\textrm{5,ln}^1 \, .
\ee
Here $\bar{d}_\textrm{3,c}^{\,1}$, $\bar{d}_\textrm{4,c}^{\,1}$, $a_\textrm{5,c}^1$, and $a_\textrm{5,ln}^1$
denote linear-in-$\nu$ subcoefficients of the coefficients
$\bar{d}_\textrm{3,c}$, $\bar{d}_\textrm{4,c}$, $a_\textrm{5,c}$, and $a_\textrm{5,ln}$, respectively.
(Beware that  $a_\textrm{5,ln}$ denotes here the coefficient of $u^5 \ln u$, while, above,  $a_\textrm{5,ln}^\textrm{I, II} $
denoted the coefficients of $r'^{-5} \ln r'= - u^5 \ln u$.)
Our analytical results yield the following value for the quantity $\rho^\textrm{c}_4$:
\be
\rho^\textrm{c ana}_4 = 64.6406.
\ee
while the numerical value obtained in Eq.\ (49) of \cite{Barack:2010ny} was
\be
\rho^\textrm{c num}_4 = 69^{+7}_{-4}.
\ee
Again the agreement is essentially at the one sigma level, which is satisfactory.

\section{Contributions to $\bar{D}(u)$ from higher PN orders: 5PN and 5.5PN}

In this section, we shall complete the above 4PN-accurate results on the second EOB potential $\bar{D}(u)$ by deriving,
using various techniques, both the 5PN and the 5.5PN contributions to $\bar{D}(u)$.
The logarithmic 5PN, as well as the 5.5PN, contributions will be derived analytically (and exactly, as functions of $\nu$)
using  effective action techniques. By contrast, only the linear-in-$\nu$ piece in the nonlogarithmic 5PN contribution
will be estimated below from the numerical  gravitational self-force computations of precession effects of Ref.\ \cite{Barack:2010ny}.
We shall then compare this PN knowledge of the linear-in-$\nu$ piece in $\bar{D}(u)$ to its numerical determination
in the strong-field domain \cite{Akcay:2012ea}.

\subsection{4PN and 5PN logarithmic contributions to the EOB dynamics}

Let us sketch in this subsection the derivation of the 4PN and 5PN logarithmic terms in the two-body Hamiltonian given by one of us \cite{Damourlogs}
and previously used to compare the EOB formalism to numerical gravitational self-force computations of precession effects \cite{Barack:2010ny}.
This derivation used a (fractionally 1PN accurate) effective action technique which might be of interest for other applications.

Let us consider the coupling of a localized self-gravitating system to an external (relativistic) tidal field.
We wish to reach a fractional 1PN accuracy in the coupling between the system and the external tidal field.
To do so it is very convenient to use the Damour-Soffel-Xu (DSX) formalism \cite{Damour:1990pi,Damour:1991yw,Damour:1992qi,Damour:1993zn}.
The DSX formalism allows one to treat the physically nonlinear effects present in 1PN self-gravity by means of simple, linear equations.
To do this one must use the following exponential parametrization of the metric \cite{Blanchet:1989ki,Damour:1990pi}:
\begin{subequations}
\label{eqA1}
\begin{align}
g_{00} &= -\exp \left( -\frac{2w}{c^2} \right) ,
\\
g_{0i} &= - \frac{4w_i}{c^3} ,
\\
g_{ij} &= \delta_{ij} \exp \left( + \frac{2w}{c^2} \right) .
\end{align}
\end{subequations}
In this decomposition $w_{\mu} = (w,w_i)$ satisfy linear equations, so that we can linearly decompose $w_{\mu}$ as $w_{\mu} = w_{\mu}^+ + \bar w_{\mu}$,
where $w_{\mu}^+$ describes the self-gravitational field (at 1PN accuracy) and $\bar w_{\mu}$ describes a general external tidal field (also treated at 1PN fractional accuracy).
It is enough to treat the coupling to the external tidal field at first order in the tidal field $\delta^{\rm ext} g_{\mu\nu}$. Therefore this coupling is described by the action
\be
\label{eqA2}
S_{\rm int} = \int \frac{\md^4 x}{c}  \, \frac12 \, \sqrt g \ T^{\mu\nu} \delta^{\rm ext} g_{\mu\nu} .
\ee
Thanks to the exponential parametrization (\ref{eqA1}) one sees that $\frac12 \, \sqrt g \ \delta^{\rm ext} g_{\mu\nu}$
is (to 1PN accuracy) linear in $\delta^{\rm ext} w_{\mu} \equiv \bar w_{\mu}$ (e.g. $\frac12 \, \sqrt g \ \delta g_{00} = \delta w / c^2$), so that we get
\be
\label{eqA3}
S_{\rm int} = \int \frac{\md^4 x}{c}  \left( \sigma \bar w - \frac4{c^2} \, \sigma^i \bar w_i \right),
\ee
where we denoted (as usual \cite{Blanchet:1989ki,Damour:1990pi})
\be
\label{eqA4}
\sigma := \frac{T^{00} + T^{ii}}{c^2}, \quad
\sigma^i := \frac{T^{0i}}c .
\ee
We can then insert in the coupling (\ref{eqA3}), which is linear in $\bar w_{\mu}$, the (linear) 1PN-accurate, general tidal expansion given by Eqs.\ (4.15) in \cite{Damour:1991yw}. The latter expansion is a linear form in the external tidal moments $G_L^{\rm ext} (t) \equiv G_{i_1 i_2 \ldots i_{\ell}}^{\rm ext}$ and $H_L^{\rm ext} (t) \equiv H_{i_1 i_2 \ldots i_{\ell}}^{\rm ext}$ and their time derivatives (we use the multi-index notation of \cite{Blanchet:1989ki,Damour:1990pi}). Operating by parts, one finds that the coupling (\ref{eqA3}) yields (modulo a boundary term arising from a total time derivative) the remarkably simple result
\begin{multline}
\label{eqA5}
S_{\rm int} = \sum_{\ell} \frac1{\ell!} \int \md t \Bigl[ G_L^{\rm ext} (t) \, M_L^{\rm BD} [y(t)] 
\\
+ \frac{\ell}{c^2 (\ell+1)} \, H_L^{\rm ext} (t) \, S_L^{\rm BD} [y(t)] \Bigl] ,
\end{multline}
where $M_L^{\rm BD} [y(t)]$, $S_L^{\rm BD} [y(t)]$ are the 1PN-accurate Blanchet-Damour (mass and spin) multipole moments of the considered localized system [symbolized as $y(t)$] whose coupling to an external (tidally expanded) field we are considering. We recall that $M_L^{\rm BD}$ and $S_L^{\rm BD}$ are certain (compact-support) spatial integrals involving various moments of the densities $\sigma$ and $\sigma^i$, Eqs.\ (\ref{eqA4}) (see \cite{Blanchet:1989ki,Damour:1990pi}). (At Newtonian order the quadrupole $M_{ab}^{\rm BD}$
reduces to the usual quadrupole moment, denoted $I_{ab}$ above.)
Note that the general action (\ref{eqA5}) can be useful for various applications. For instance, it can be used to describe 1PN-accurate radiation-reaction effects. Indeed, Refs.\ \cite{Blanchet:1984wm,Blanchet:1996vx} (generalizing Ref.\ \cite{Burke:1970wx}) have shown that radiation reaction can be described by external-like potentials $w_{\mu}^{\rm RR}$, which are parametrized by tidal moments given by time derivatives of the 1PN-accurate Blanchet-Damour moments $M_L^{\rm BD}$, $S_L^{\rm BD}$, when working at the 1PN fractional accuracy \cite{Blanchet:1996vx}, namely
\begin{subequations}
\label{eqA6}
\begin{align}
G_{ab}^{\rm RR \, (1PN)} (t) &= -\frac25 \, \frac G{c^5} \, M_{ab}^{{\rm BD} (5)} ,
\\[1ex]
G_{abc}^{\rm RR} (t) &= +\frac2{63} \, \frac G{c^7} \, M_{abc}^{{\rm BD} (7)} ,
\\[1ex]
H_{ab}^{\rm RR} (t) &= -\frac{16}{15} \, \frac{G}{c^5} \, S_{ab}^{\rm BD(5)} ,
\end{align}
\end{subequations}
where the superscript, say in $M_L^{{\rm BD}(n)}$ denotes the $n$th time derivative. One can check that the coupling defined by (\ref{eqA5}),
(\ref{eqA6}) implies [when varying only the system variables $y(t)$ in (\ref{eqA5})] the 1PN-accurate radiation-reaction equations of motion of Ref.\ \cite{Iyer:1995rn}.

Moreover, the same result (\ref{eqA5}), (\ref{eqA6}) can be used to compute the {\it conservative} logarithmic contributions arising at 4PN and 5PN.
Indeed, Ref.\ \cite{Blanchet:1987wq} has shown that the first logarithmic contribution arose at 4PN from the fact that the (near zone) radiation reaction quadrupole $G_{ab}^{\rm RR} (t)$ was the fifth derivative of a tail-modified quadrupole moment $M_{ab}^{\rm tail}$ involving an integral over the past of the type (for $\ell = 2$)
\begin{multline}
\label{eqA7}
M_L^{\rm RR} (t) = M_L (t) + \frac{4 G {\mathcal M}}{c^3} \int_0^{+\infty} \md\tau M_L^{(2)} (t-\tau) 
\\[1ex]
\times \left[ \ln \left( \frac{c\tau}{2 r_0} \right) + \kappa_{\ell}^{\rm RR} \right]
\end{multline}
[and similarly for $S_L^{\rm RR} (t)$], where ${\mathcal M}$ denotes the total, Arnowitt-Deser-Misner mass of the local system (henceforth treated as a nondynamical constant),
and where $r_0 $ is the length scale entering  the Blanchet-Damour multipolar-post-Minkowskian formalism \cite{Blanchet:1985sp}. Note that the coefficient of the integral (tail) term is $4G {\mathcal M} / c^3$, i.e.\ {\it twice} the coefficient of the tail term entering the (wave zone) radiative multipole moments \cite{Blanchet:1989ki,Blanchet:1992br,Blanchet:1995fr}. The reason for this factor 2 is clearly general \cite{Blanchet:1992br} so that the structure (\ref{eqA7}) applies for all values of $\ell$ (and for both types of moments). As a consequence, the $\ln (1/c)$ contribution to $M_L^{\rm RR}$ is given by
\be
\label{eqA8}
M_L^{\rm RR}  (t) = M_L (t) - 4 \, \frac{G{\mathcal M}}{c^3} \, \ln \left( \frac1c \right) M_L^{(1)} (t) .
\ee

The corresponding $\ln\frac1c$ contribution to the equations of motion is found to be conservative
(though it comes from a tail modification of a dissipative effect) \cite{Blanchet:1987wq}.
As a consequence of the {\it bilinear} structure of the action (\ref{eqA5}),
the 4PN and 5PN conservative logarithmic effects can be very simply obtained by:
(i) inserting the logarithmic contribution of (\ref{eqA8}) in (\ref{eqA6}) and then in (\ref{eqA5})
(with $G_L^{\rm ext} = G_L^{{\rm RR} \ln}$, $H_L^{\rm ext} = H_L^{{\rm RR} \ln}$);
(ii) operating by parts to equalize the number of derivatives on $M_L$ and $S_L$;
and, last but not least, (iii) by adding an overall factor $\frac12$
[because only $M_L [y(t)]$, $S_L [y(t)]$ were supposed to be varied in (\ref{eqA5}),
while we now wish to have an action where one has to vary two occurrences of $M_L$ in a bilinear form $\sum G_L [M_L] M_L$].
This yields the simple action
\be
\label{eqA8bis}
S^{\ln c} [y] = -2 \, \frac{G {\mathcal M}}{c^3} \ln \left(\frac1c\right) \int \md t\, {\mathcal F}_E^{\rm 1PN} [y(t)],
\ee
where
\begin{align}
\label{eqA9}
{\mathcal F}_{E}^{\rm 1PN}[y] &= \frac G{c^5} \bigg( \frac15 \, (M_{ab}^{\rm BD(3)})^2 + \frac1{189 c^2} (M_{abc}^{\rm BD(4)})^2 
\nonumber\\[1ex]&\quad
+ \frac{16}{45 c^2} (S_{ab}^{\rm BD(3)})^2 \bigg)
\end{align}
denotes the fractionally 1PN-accurate instantaneous energy flux radiated as gravitational waves by the system
[considered as a function of the dynamical variables $y\sim(\vecr_a,\vecp_a)$ describing the localized system, say a two-body system].
The corresponding Hamiltonian is simply
\be
\label{eqA10}
H_{\rm 4PN + 5PN}^{\ln c} = + \, \frac{2G{\mathcal M}}{c^3} \ln \left( \frac1c \right) {\mathcal F}_E^{\rm 1PN} .
\ee
This remarkably simple result generalizes the 4PN logarithmic contribution of the Hamiltonian of Ref.\ \cite{Damour:2014jta}
[seen from the near zone, the $\ln r$ term is a $\sim \ln (\Omega r/c) \sim \ln \left( \frac1c \right)$ contribution].
It suggests that the nonlocal part $\hat H^{\rm nr \, II}$, Eq.\ (\ref{hrii}), can be simply generalized to
\be
\label{eqA11}
- \frac{G{\mathcal M}}{c^3} \, {\rm Pf}_{2 s_{\rm phys} / c} \int_{-\infty}^{+\infty} \frac{\md\tau}{\vert \tau \vert} \,
{\mathcal F}_E^{\rm split} (t,t+\tau),
\ee
where ${\mathcal F}_E^{\rm split} (t,t')$ is obtained from (\ref{eqA9}) by splitting each squared multipole
according to $(M_L^{(\ell+1)})^2 \to M_L^{(\ell + 1)} (t) \, M_L^{(\ell + 1)} (t')$.

The result (\ref{eqA10}) yields also a straightforward way to compute the logarithmic contributions to the EOB Hamiltonian at the 4PN and 5PN levels. A simple way to do it is to use the strategy we used in the text for $\hat H^{\rm nr \, II}$, namely to compute separately the ($\ell$-average) Delaunay Hamiltonian corresponding to (\ref{eqA10}), and the $\ell$-averaged squared effective Hamiltonian, and to relate their logarithmic contributions via
\be
\label{eqA12}
\delta^{\ln} \left\langle \hat H_{\rm eff}^2 \right\rangle_{\ell} = \frac4\nu \, \frac G{c^3} \ln \left( \frac1c \right) \hat H_{\rm eff} \left( \frac H{(m_1+m_2) c^2} \right)^2 \langle {\mathcal F}_E^{\rm 1PN} \rangle_{\ell} .
\ee
Note that the $\ell$-averages (here indicated by $\langle \ldots \rangle_{\ell}$) have to be done at the 1PN accuracy (which is most easily done by using the 1PN quasi-Keplerian parametrizations of \cite{DamourDeruelleAIHP1985}). After straightforward calculations, starting from the expression of ${\mathcal F}_E^{\rm 1PN}$ given in \cite{Blanchet:1989cu},
one finds that the 4PN and 5PN logarithmic contributions to $A(u)$, $\bar D(u)$ and $\hat Q(u,p_r)$ are (using the replacement $\ln \frac1c \equiv \frac12 \ln \frac1{c^2} \to \ln u$; as in \cite{Damour:2009sm})
\begin{subequations}
\label{eqA13}
\begin{align}
A^{\ln} (u) &= \frac{64}5 \nu u^5 \ln u + \left(-\frac{7004}{105} \nu - \frac{144}5 \nu^2 \right) u^6 \ln u ,
\\[1ex]
\bar D^{\ln} (u) &= \frac{592}{15} \nu u^4 \ln u + \left( -\frac{1420}{7} \nu - \frac{3392}{15} \nu^2 \right) u^5 \ln u ,
\\[1ex]
\hat Q^{\ln} (u,p_r) &= \left( \frac{5428}{105} \nu - \frac{592}{5} \nu^2 \right) u^4 p^4_r \ln u .
\end{align}
\end{subequations}
The 4PN contributions ($\sim u^5 \ln u$ in $A$ and $\sim u^4 \ln u$ in $\bar D$) agree with our complete 4PN results above,
while the other terms are additional 5PN contributions to the EOB formalism.
At the time of their derivation (see \cite{Damourlogs}) a comparison with the fractionally 1PN-accurate computation
by Blanchet \emph{et al.\@} \cite{Blanchet:2010zd} of the energetics of circular binaries [entirely encoded in the $A(u)$ function]
showed agreement with $A^{\ln} (u)$ above for the terms linear in $\nu$, but disagreement for the coefficient of $\nu^2 u^6 \ln u$.
However, the later correction, mentioned in \cite{LeTiec:2011ab}, to the results of \cite{Blanchet:2010zd} yielded perfect agreement with $A^{\ln} (u)$ above
[as independently derived in \cite{Barausse:2011dq}, which also deduced part of the $D^{\ln} (u)$ above from the combined results published in \cite{Barack:2010ny}].

\subsection{5.5PN contributions to the EOB dynamics}

It has been recently realized that there existed half-integral-order PN contributions to the {\it conservative} dynamics of binary systems
\cite{Shah:2013uya,Bini:2013rfa,Blanchet:2013txa,Blanchet:2014bza}.  As explained in \cite{Bini:2013rfa}, such terms (which seem to violate
the usual PN lore that conservative effects are associated with even powers of $1/c$) have the same conceptual origin as the logarithmic
5PN contributions discussed above; namely the fact  that the tail-transported hereditary contribution to the near-zone gravitational field \cite{Blanchet:1987wq}
is time dissymmetric, without being time antisymmetric. We shall use the same effective-action approach as in the previous subsection but extend it 
(as in \cite{Bini:2013rfa}) to the second-order tail contribution.  The relevant term in the near-zone metric is
[see Eq.\ (79) in \cite{Bini:2013rfa}, or, after time symmetrization, Eq.\ (5.13a) in \cite{Blanchet:2013txa}]
\begin{multline}
\label{tail2metric}
g_{00}^{{\rm tail}^2} (t,\mathbf{x}) =
- \frac{2G}{5c^7} x^i x^j 4 B \left(\frac{G \mathcal{ M}}{c^3} \right)^2
\\[1ex]
\times \int_0^{+\infty} \md\tau M_{ij}^ {(8)} (t-\tau)   \ln\left(\frac{c \tau}{2 r'_0}\right),
\end{multline}
where 
\be
B = -\frac{107}{105},
\ee
and where $r'_0=r_0\exp(-11/12)$ is a rescaled version of the length scale
entering the multipolar-post-Minkowskian formalism of Ref.\ \cite{Blanchet:1985sp}.
(Actually, the precise value of $r'_0 $ is unimportant, as the scale dependence will disappear after taking the time symmetrization.)

Starting from Eq.\ (\ref{tail2metric}), one must time symmetrize it,
and then insert it into the effective action [see Eq.\ (\ref{eqA3})]
\be
\label{eqB1}
S_{\rm int} = \int \md t \, \frac12 \sum_a m_a c^2 \, g_{00}^{{\rm tail}^2} .
\ee

As above, the time differentiated quadrupole moment $M_{ij}^{(8)}$ entering this action must be considered, at this stage, as an {\it external} tidal moment.
However, after the time symmetrization (and after a convenient integration by parts over the $\tau$ integral) one finds that $S_{\rm int}$ involves the double integral
\be
\label{eqB2}
\int \md t \int_{-\infty}^{+\infty} \frac{\md\tau}{\tau} \,
M_{ij} (t) \left[ M_{ij}^{{\rm ext} (7)} (t-\tau) - M_{ij}^{{\rm ext} (7)} (t+\tau) \right],
\ee
where the $\tau$ integral has been extended to the full range $-\infty , + \infty$ (which simply doubles the value of the integral).

When integrating seven times by parts the $t$ integral, one sees that the bilinear form (\ref{eqB2}) is (despite first appearances) {\it symmetric}
under the exchange between the two functions $M_{ij} (t)$ and $M_{ij}^{\rm ext} (t')$.
(One works modulo total time derivatives, as appropriate for an action density.)
As in the previous subsection, one can then obtain an autonomous action for the corresponding (conservative) dynamics,
simply by taking $M_{ij}^{\rm ext} = M_{ij}$, and including an extra factor $\frac12$.

Finally, one ends up with an action which (after some operations by parts) can be written as
\begin{multline}
\label{eqB3}
S^{{\rm tail}^2} = -\frac{B}{10}\,\frac{G}{c^5} \left( \frac{G {\mathcal M}}{c^3} \right)^2
\int \md t \int_{-\infty}^{+\infty} \frac{\md\tau}{\tau} \, M_{ij}^{(3)} (t) 
\\[1ex]
\times \left[ M_{ij}^{(4)} (t+\tau) - M_{ij}^{(4)} (t-\tau) \right].
\end{multline}
It could alternatively be expressed in terms of
\begin{multline}
\label{eqB4}
\int \md t \int_{-\infty}^{+\infty} \md\tau \ln \left(\frac{\vert\tau\vert}{\tau_0}\right) M_{ij}^{(4)} (t)
\\[1ex]
\times \left[ M_{ij}^{(4)} (t+\tau) + M_{ij}^{(4)} (t-\tau) \right]
\end{multline}
(whose time symmetry is more evident), but the form (\ref{eqB3}) has the advantage of clearly exhibiting
the independence on the choice of any logarithmic time scale $\tau_0=2r'_0/c$.

Inserting in Eq.\ (\ref{eqB3}) the Fourier expansion (\ref{IBessel}) of the quadrupole moment [$M_{ij} = I_{ij} + {\mathcal O}(1/c^2)$]
leads to a series of integrals of the form
\be
\label{eqB5}
\int_{-\infty}^{+\infty} \frac{\md\tau}{\tau} \sin(p\Omega\tau)
= {\rm sign} (p) \int_{-\infty}^{+\infty} \frac{\md x}{x} \sin x
= \pi \,{\rm sign} (p),
\ee
where ${\rm sign} (p)$ is defined as being $+1$ if $p > 0$, $-1$ if $p<0$, and $0$ if $p=0$.
Finally the {\it averaged} second-order tail Hamiltonian is easily found to be given by
\be
\label{eqB6}
\bar H^{{\rm tail}^2} = - \pi \, \frac{2B}{5} \, \frac{G}{c^5} \left( \frac{G {\mathcal M}}{c^3} \right)^2 \Omega_{\rm phys}^7 \, S_7^{\rm quad},
\ee
where $ \Omega_{\rm phys}= \Omega/(GM)$ and 
\be
\label{eqB7}
S_7^{\rm quad} := \sum_{p=1}^{\infty} p^7 \vert I_{ij} (p) \vert^2 .
\ee

While the averaged 4PN Hamiltonian involved series of the type [see Eqs.\ (\ref{Faverage}) and (\ref{intcos})]
\be
\label{eqB7bis}
\sum_{p=1}^{\infty} p^6 (a \ln p+b) \, \vert I_{ij} (p) \vert^2 ,
\ee
we see that the second-order tail Hamiltonian involves the series (\ref{eqB7}) containing the seventh powers of $p$
[or, more completely, in view of Eq.\ (\ref{eqB5}), the seventh powers of the absolute values of all the relative integers $p$ entering the expansion (\ref{IBessel})].
The series (\ref{eqB7}) entered the calculation by Arun \emph{et al.\@} of the (first-order) tail contribution to the averaged gravitational-wave energy flux \cite{Arun:2007rg}.
Using the definition (5.4) of Ref.\ \cite{Arun:2007rg}, one has
\be
\label{eqB8}
S_7^{\rm quad} = \mu^2 (GM a)^4 \, 32 \, \varphi(e),
\ee
where the normalized, dimensionless eccentricity function $\varphi (e)$ can be expanded in powers of $e^2$
with the result \cite{Arun:2007rg,Arun:2009mc}
\be
\label{eqB9}
\varphi (e) = 1 + \frac{2335}{192} \, e^2 + \frac{42955}{768} \, e^4 + {\mathcal O} (e^6) .
\ee

Inserting Eq.\ (\ref{eqB8}) into Eq.\ (\ref{eqB6}) (in which it is enough to use the approximations ${\mathcal M} \simeq M$ and $GM \Omega_{\rm phys} = \Omega \simeq a^{-3/2}$) yields
\be
\label{eqB10}
\frac{\bar H^{{\rm tail}^2}}{M c^2} = \frac12 \, C \, \frac{\nu^2}{(c^2 a)^{6.5}} \, \varphi (e) ,
\ee
where we recall that $a=a_{\rm phys}/(GM)$, and where
\be
\label{eqB11}
C = - \frac{128}{5} \, B \, \pi = + \frac{13696}{525} \, \pi .
\ee
The averaged Hamiltonian (\ref{eqB10}) is to be considered as a function of the action variables ${\mathcal L}$ and ${\mathcal G}$,
via Eqs.\ (\ref{ae-vs-LG}).

Using the technique explained above, we can transcribe (\ref{eqB10}) within the EOB formalism
by considering corresponding additional (tail-of-tail) 5.5PN contributions to the EOB potentials of the form
\begin{subequations}
\label{eqB12}
\begin{align}
A(u) &= 1 - 2 u + \cdots + A_{6.5} (\nu) \, u^{6.5} ,
\\[1ex]
\bar D (u) &= 1 + \cdots + \bar D_{5.5} (\nu) \, u^{5.5} ,
\\[1ex]
\hat Q_{5.5} (u,p_r) &= q_{44.5} (\nu) \, p_r^4 \, u^{4.5} + q_{63.5} (\nu) \, p_r^6 \, u^{3.5} + \mathcal{O}(p_r^8) .
\end{align}
\end{subequations}
The matching between (\ref{eqB10}) and the effective EOB Hamiltonian goes via (setting $c=1$)
\begin{multline}
\label{eqB13}
\frac2\nu \, \frac{\bar H^{{\rm tail}^2}}{M} = \left\langle \delta^{{\rm tail}^2} \, \hat H_{\rm eff}^2 \right\rangle_{\!\ell} 
\\[1ex]
= \left\langle A_{6.5} \, u^{6.5} + \bar D_{5.5} \, u^{5.5} \, p_r^2 + \hat Q_{5.5} (u,p_r) \right\rangle_{\!\ell} .
\end{multline}

The $\ell$-averages on the right-hand side can be done at the Newtonian level.
A convenient way of evaluating them is to transform them into $f$-averages,
where $f$ denotes the true anomaly [see Eq.\ (\ref{eq4.12})], using, for any dynamical function $F$
\be
\label{eqB14}
\langle F \rangle_{\ell} = \frac1{\sqrt{1-e^2}} \, \left\langle \left( \frac ra \right)^2 F \right\rangle_{\!\!f} .
\ee
As indicated in Eq.\ (\ref{eqB12}) the 5.5PN contribution to $\hat Q(u,p_r)$ formally contains an infinite string of powers of $p_r$,
starting at $\mathcal{O}(p_r^4)$. There is a one-to-one correspondence between $\{ A_{6.5} , \bar D_{5.5} , q_{44.5} , q_{63.5} , \ldots \}$
and the infinite list of Taylor coefficients of the eccentricity expansion of $C\,\varphi(e)=C(1 + \varphi_2 \, e^2 + \varphi_4 \, e^4 + \cdots)$.
At order $e^0$ we see that
\be
\label{eqB15}
A_{6.5} (\nu) = C \nu = \frac{13696}{525} \, \pi \, \nu
\ee 
in agreement with the results of Ref.\ \cite{Bini:2013rfa} (see also Refs.\ \cite{Shah:2013uya,Blanchet:2013txa}).
We get the new result we were looking for by identifying the $\mathcal{O}(e^2)$ contribution to the right-hand side of (\ref{eqB13})
to the corresponding term $\propto C \, \varphi_2$ in Eq.\ (\ref{eqB10})
[using the $\mathcal{O}(e^2)$ term \cite{Arun:2007rg} in Eq.\ (\ref{eqB9})]. This yields
\be
\label{eqB16}
\frac{\bar D_{5.5}}{A_{6.5}} = \frac{619}{96},
\ee 
i.e.
\be
\label{eqB17}
\bar D_{5.5} (\nu) = \frac{264932}{1575} \, \pi \, \nu \cong 528.4497936 \, \nu .
\ee 
Note that both $A_{6.5} (\nu)$ and $\bar D_{5.5} (\nu)$ are proportional to $\nu$. This is a useful information, especially as gravitational self-force techniques can, currently, only give access to corrections linear in $\nu$.

\subsection{Linear-in-$\nu$ piece in $\bar D(u)$: 5PN contribution and global fits}

In this subsection we consider the linear-in-$\nu$ piece (or first-order self-force piece)
in the expansion of the EOB potential $\bar D (u,\nu)$ in powers of $\nu$:
\be
\label{eqC1}
\bar D (u) = 1 + \nu \, \bar d (u) + \nu^2 \, \bar d_2 (u) + \mathcal{O}(\nu^3).
\ee
For convenience, we have omitted to put a subscript $1$ on the $\mathcal{O}(\nu)$ piece $\bar d(u)$.
Before focusing on $\bar d (u)$, let us note that our results above have given access to the first two terms in the PN expansion of the $\mathcal{O}(\nu^2)$ piece $\bar d_2 (u)$, namely
\begin{align}
\label{eqC2}
\bar d_2 (u) &= -6 \, u^3 + \left( \frac{123 \, \pi^2}{16} - 260 \right) u^4 + \mathcal{O}(u^5)
\nonumber \\[1ex]
&= -6 \, u^3 - 184.1274 \, u^4 + \mathcal{O}(u^5).
\end{align}
We note that both terms are negative, and will therefore tend to reduce the value of $\bar D (u,\nu)$ when considering comparable-mass systems.
Similarly to the discussion in \cite{Akcay:2012ea} of the behavior of the successive terms in the $\nu$-expansion of
\be
\label{eqC3}
A(u,\nu) = 1-2u + \nu a(u) + \nu^2 a_2 (u) + \mathcal{O}(\nu^3),
\ee
(see also Refs.\ \cite{Bini:2014ica,Bini:2014zxa} for a discussion of other $\nu$-expansions)
one expects that the individual contributions $\bar d (u)$, $\bar d_2(u)$, \ldots will have formal singularities at the light ring (i.e.\ when $u \to \frac13$).
Previous examples suggest that the formal singularity in $\bar d_2 (u)$ will be stronger, and of the opposite sign, than that in $\bar d(u)$.
This suggests that the negative signs in Eq.\ (\ref{eqC2}) are related to, and indicative of, a singular behavior $d_2 (u) \to -\infty$ as $u \to \frac13$.

Gravitational self-force studies of the precession of small mass-ratio orbits
have allowed one to numerically compute the value of $\bar d(u)$
on a sample of orbits spanning the interval $0.0125 \leq u \leq \frac16$ \cite{Barack:2010ny,Akcay:2012ea}.
As we have seen above, these numerical data had led to a numerical estimate of the nonlogarithmic 4PN coefficient
say $\bar d_4^c$ (denoted $\bar d_{4,c}^1$ above) which is consistent with our 4PN result.
On the other hand, the determination of the higher-order PN coefficients,
and notably of the nonlogarithmic 5PN coefficient $\bar d_5^c$, was extremely coarse.
Namely, the transcription \cite{Barausse:2011dq} of the result $-6000 \leq \rho_r^c \leq -440$ [Eq.\ (49) in \cite{Barack:2010ny}] yielded
\be
\label{eqC4}
-1849 \leq \bar d_5^c \leq -249 .
\ee

The additional knowledge we have acquired above about the analytical value of $\bar d_4^c$, as well as that of the 5.5PN term $\bar d_{5.5} = \bar D_{5.5} (\nu) / \nu$,
opens the opportunity of extracting more PN information from the numerical data on the function $\bar d(u)$ presented in Table~VII of \cite{Akcay:2012ea}.
We explored various PN-like fits of the latter data on $\bar d(u)$, keeping fixed all analytically known PN coefficients,
and fitting for some higher-order PN expansion parameters involving a nonlogarithmic 5PN term $\bar d_5^c \, u^5$,
as well as a combination of terms of order $u^6$ and/or $u^6 \ln u$ and/or $u^{6.5}$ and/or $u^7$.
In order to avoid the contamination of higher-order PN terms (which become important in the strong-field domain),
we have first tried to determine the value of $\bar d_5^c$ from {\it local} fits, namely (error-weighted least-squares) fits on the weak-field interval $0.0125 \leq u \leq 0.1$.
We explored fits involving, besides $\bar d_5^c$, either one, two or sometimes three extra parameters.
From the range of values of $\bar d_5^c$ resulting from such fits, we concluded that the value of $\bar d_5^c$ is probably in the range
\be
\label{eqC5}
\bar d_5^{c \, {\rm num}} = - 343 \pm 76 .
\ee
This range significantly improves the previous knowledge, Eq.\ (\ref{eqC4}), on $\bar d_5^c$.

\begin{figure}[h]
$$
\includegraphics[height=8cm]{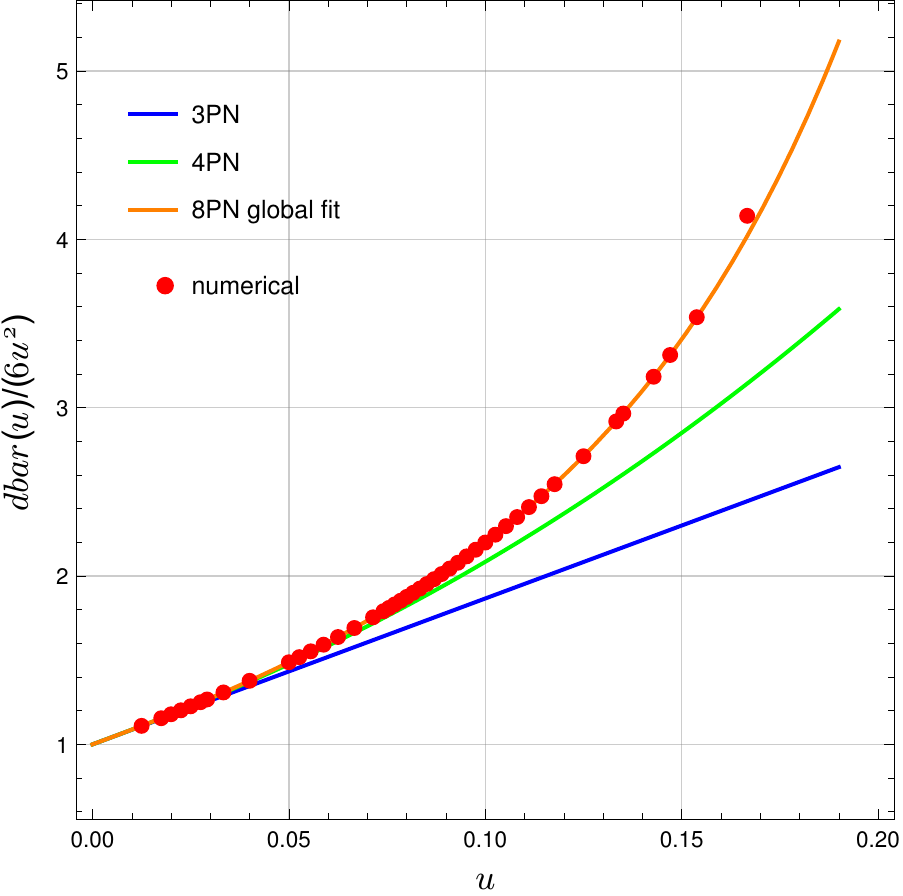}
$$
\caption{Comparison of the numerical result \cite{Akcay:2012ea} $\bar{d}^{\,\rm num}(u)/(6u^2)$ for the rescaled function $\bar{d}(u)/(6u^2)$
with the global fit $\bar{d}^{\,\rm glob}(u)/(6u^2)$,  Eq.~\eqref{eqC7}, and the analytically known 3PN \cite{Damour:2000we} and 4PN results (derived above).}
\label{fig:1}       
\end{figure}

We also performed some {\it global} PN-like fits of $\bar d^{\rm num} (u)$ on the full available interval $0.0125 \leq u \leq \frac16$.
To reach a decent goodness of fit on such a large interval, one needs to include at least four parameters.
For instance, we found that a weighted least-squares fit involving $u^5$, $u^6$, $u^7$, and $u^8$
gave a reduced $\chi^2$ equal to $2.14$ for the following coefficients:
\begin{align}
\label{eqC6}
\bar d_5^{ \, c \, {\rm glob}} &= -396.905 ,
\nonumber \\
\bar d_6^{ \ {\rm glob}} &= 4935.68 ,
\nonumber \\
\bar d_7^{ \ {\rm glob}} &= -17634.7 ,
\nonumber \\
\bar d_8^{ \ {\rm glob}} &= 134021 .
\end{align}
In other words, the PN-like expansion
\begin{align}
\label{eqC7}
\bar d^{ \, {\rm glob}} (u) &= 6 \, u^2 + 52 \, u^3 + \left( 221.5719912 + \frac{592}{15} \ln u \right) u^4
\nonumber \\&\quad
+ (\bar d_5^{ \, c \, {\rm glob}} - 202.8571429 \ln u) \, u^5
\nonumber \\[1ex]&\quad
+ 528.4497936 \, u^{5.5} + \bar d_6^{ \ {\rm glob}} \, u^6
\nonumber \\[1ex]&\quad
+ \bar d_7^{ \ {\rm glob}} \, u^7 + \bar d_8^{ \ {\rm glob}} \, u^8 ,
\end{align}
which combines our analytical results with the numerical result (\ref{eqC6}), yields a reasonably accurate representation of $\bar d(u)$ on the full interval $0 \leq u \leq \frac16$.
In Fig.\ \ref{fig:1} we compare the rescaled function $\bar d^{ \, {\rm glob}} (u) / (6 \, u^2) = 1 + \frac{26}3 \, u + \cdots$
to the corresponding numerical result $\bar d^{ \, \rm num} (u) / (6 \, u^2)$ (which was plotted in Fig.\ 8 of \cite{Akcay:2012ea}).
We also show, for comparison, some of the analytically known 3PN (i.e.\ $1+\frac{26}3 \, u$) and 4PN corresponding results.

Further studies are called for to see whether the new knowledge on the function $\bar D (u,\nu)$,
namely [combining Eqs.\ (\ref{eqC2}) and (\ref{eqC7})],
\be
\label{eqC8}
\bar D (u,\nu) \simeq 1 + \nu \, \bar d^{ \, \rm glob} (u) + \nu^2 \, \bar d_2 (u) ,
\ee
can be incorporated to improve the current EOB models.
[Until now, one has only used the 3PN approximation to $\bar D (u,\nu) = 1 + \bar D_2 (\nu) \, u^2 + \bar D_3 (\nu) \, u^3$.]
As was the case for the main EOB radial potential $A(u,\nu)$, it might become necessary to replace the PN-like expansion of $\bar D (u,\nu)$ by some resummed one.
And as was the case for $A(u,\nu)$, it might then be necessary to appeal to specific numerical simulations to calibrate such a resummed form of $\bar D (u,\nu)$.
There are several numerical experiments that might inform the strong-field values of $\bar D (u,\nu)$,
notably periastron advance studies (such as \cite{LeTiec:2011bk}) and strong-field scattering simulations (such as \cite{Damour:2014afa}).

\section{Conclusions}

Let us summarize our main results.

We decomposed the recently derived, nonlocal-in-time 4PN Hamiltonian \cite{Damour:2014jta} into two parts:
(i) a local-in-time part $H^{\rm I}$, Eq.\ (\ref{eq2.3bis}),
[which includes an explicit logarithmic contribution $F(\vecr,\vecp)\ln(r/s)$, where $s$ is a scale separating near-zone from wave-zone effects];
and (ii) a nonlocal-in-time part $H^{\rm II}$, Eq.\ (\ref{hrii}), (which involves an integration over arbitrarily large time separations).

We transcribed the local dynamics $H^{\rm I}$ into an equivalent (gauge-invariant) EOB description
by performing a 4PN-accurate canonical transformation between the Arnowitt-Deser-Misner phase space and the EOB one.
This led to EOB potentials $A^{\rm I} (r')$, $\bar D^{\rm I} (r')$ and $Q^{\rm I}(r',p'_r)$ given (in exact form) by Eqs.\ (\ref{eobAD2}), (\ref{eobAD5}), (\ref{eobQ2}) with Eqs.\ (\ref{match1res}).
Note, in particular, that $Q^{\rm I}$ is a polynomial in $p'_r$ which involves only $p'^4_r$ and $p'^6_r$.

On the other hand, we had to introduce a new technique for transcribing the nonlocal part $H^{\rm II}$
into an equivalent EOB description by means of ordinary (local-in-time) potentials $A^{\rm II} (r')$, $\bar D^{\rm II} (r')$ and $Q^{\rm II}(r',p'_r)$. Our technique combined several steps:
(i) a formal order reduction allowing one to express the time-separated phase-space variables $\vecr (t+\tau)$, $\vecp (t+\tau)$ as functions of $\vecr (t)$, $\vecp (t)$;
(ii) a transformation to action-angle variables $\ell , {\mathcal L} ; g , {\mathcal G}$;
and, finally (and crucially) (iii) the use of Delaunay's method, i.e.\ the elimination of periodic terms in the angle variables by suitable canonical transformations.
After formally making an infinite number of such Delaunay transformations,
we ended up with a new Delaunay-like Hamiltonian $\bar H^{\rm II}({\mathcal L},{\mathcal G})$
depending only on the action variables ${\mathcal L}$, ${\mathcal G}$ (and given by an angle average of the original $H^{\rm II}$).
(Let us note in passing that the original derivation of the 2PN EOB dynamics \cite{Buonanno:1998gg} was similarly based on the 2PN-accurate Delaunay Hamiltonian derived in \cite{Damour:1988mr}.)
The resulting Delaunay Hamiltonian $\bar H^{\rm II} ({\mathcal L} , {\mathcal G})$ cannot be translated into explicit, polynomial EOB counterparts $A^{\rm II} , \bar D^{\rm II} , Q^{\rm II}$.
Indeed, the last potential $Q^{\rm II} (r' , p'_r)$ must be taken as an infinite series in $p'_r$, Eq.\ (\ref{QIIfull}),
corresponding to the infinite expansion, Eq.\ (\ref{eq7.4}), expressing $\bar H^{\rm II} ({\mathcal L} , {\mathcal G})$ in terms of the Fourier-Bessel expansion, Eq.\ (\ref{IBessel}),
of the quadrupole moment of the system. Here, we have determined the value of $Q^{\rm II}$ up to $p'^6_r$ (i.e.\ the maximum power of $p'_r$ entering its first part $Q^{\rm I}$).
Our results for $A^{\rm II}$, $\bar D^{\rm II}$ and $Q^{\rm II}$ are given in Eqs.\ (\ref{match2res}).

By combining parts I and II, we could transcribe [modulo $\mathcal{O}(p'^8_r)$ terms] the original nonlocal Hamiltonian
into equivalent 4PN-accurate EOB potentials $A$, $\bar D$ and $Q$, see Eqs.\ (\ref{eqs8.1}).
We checked that the nonlogarithmic 4PN contribution to $A$ and $\bar D$ agree with previous (analytical or numerical) results.

To complete our 4PN-level results, we used an effective action technique to compute some higher-order contributions.
More precisely, we showed how to directly derive the logarithmic contributions to the EOB dynamics arising at 4PN and 5PN, Eqs.\ (\ref{eqA13}),
as well as the half-integral contributions first showing up at the 5.5PN level [Eqs.\ (\ref{eqB15}) and (\ref{eqB17})]. 

Finally, we used this knowledge to improve the extraction of PN information from gravitational self-force numerical data on the linear-in-$\nu$ piece $\bar d (u)$ in $\bar D (u,\nu)$.
This led to an approximate determination of the nonlogarithmic 5PN contribution to $\bar d (u)$, Eq.\ (\ref{eqC5}),
and to a simple, PN-like representation of the behavior of $\bar d (u)$ in the strong-field domain, Eqs.\ (\ref{eqC7}) and (\ref{eqC8}).

\begin{acknowledgments}

P.J.\ thanks the Institut des Hautes \'Etudes Scientifiques
for hospitality during a crucial stage of the collaboration.
The work of P.J.\ was supported in part by the Polish NCN grant
\textit{Networking and R\&D for the Einstein Telescope}.

\end{acknowledgments}

\end{document}